\documentclass[prb,twocolumn, superscriptaddress]{revtex4}
\usepackage{graphicx,amsmath,amssymb,amsxtra}
\renewcommand{\narrowtext}{\begin{multicols}{2} \global\columnwidth20.5pc}

\def\be{\begin{eqnarray}}
\def\ee{\end{eqnarray}}

\newcommand{\Eq}[1]{Eq.~(\ref{#1})}
\renewcommand{\o}[1]{{#1}^{\circ}}
\newcommand{\Fig}[1]{Fig.~(\ref{#1})}

\newcommand{\supp}{\operatorname{supp}}
\newcommand{\free}{\operatorname{free}}

\newcommand{\cC}{{\cal C}}
\newcommand{\cH}{{\cal H}}
\newcommand{\cR}{{\cal R}}
\newcommand{\cB}{{\cal B}}
\newcommand{\cL}{{\cal L}}
\newcommand{\cS}{{\cal S}}
\newcommand{\cN}{{\cal N}}
\newcommand{\cD}{{\cal D}}
\newcommand{\ea}{\'e }

\newcommand{\ket}[1]{{|#1\rangle}}

\begin{document}
%\draft

\title{
Linear independence of  %spin-$\frac 12$ 
nearest neighbor valence bond states on the kagom\ea
lattice and construction of SU(2)-invariant spin-$\frac12$-Hamiltonian with a
Sutherland-Rokhsar-Kivelson quantum liquid  
ground state
}

\author{Alexander Seidel}
\affiliation{
Department of Physics and Center for Materials Innovation, 
Washington University, St. Louis, MO 63136, USA}
\affiliation{
Kavli Institue for Theoretical Physics, Santa Barbara, CA 93106}
\date{\today}

\begin{abstract}

A class of local $SU(2)$-invariant spin-$\frac 12$ Hamiltonians 
is studied that has ground states within the space of
nearest neighbor valence bond states on the kagom\ea lattice.
Cases  include ``generalized Klein'' models without
obvious non-valence bond ground states, as well as a ``resonating valence-bond'' 
Hamiltonian whose unique ground states within the nearest neighbor
valence bond space are four topologically degenerate ``Sutherland-Rokhsar-Kivelson''
(SRK) type wavefunctions, which are expected to describe a gapped $\mathbb{Z}_2$
spin liquid. The proof of this uniqueness is intimately related to the linear
independence of the nearest neighbor valence bond states
on quite general and arbitrarily large kagom\ea lattices, which is also established in this work.
It is argued that the SRK ground states are also unique
within the entire Hilbert space, depending on properties of the generalized
Klein models. Applications of the strategies developed in this work to
other lattice types are also discussed.
\end{abstract}

\maketitle
\section{Introduction} 

%Anderson, Rokhsar-Kivelson, Moessner-Sondhi, Klein, Kivelson-Chayes-Chayes, Misguich, and all that.

\subsubsection{General Motivation}

For over 70 years, the study of magnetism has played a pivotal role
in defining paradigms in condensed matter physics.
% as a whole,
%and strongly correlated electron physics in particular.
Heisenberg's original notion of a local exchange interaction 
and its generalizations have given rise to a
rich and interesting 
class of Hamiltonians, whose exploration has lead to great success
both in understanding known magnetic phenomena, as well as
anticipating new ones. 
This approach has been successful 
to such a degree
that sometimes even models 
studied out of purely academic interest have 
shed light on later observed experimental
phenomena.
Prominent examples include the one-dimensional (1D) Heisenberg chain with antiferromagnetic
exchange constant, which was found to be exactly solvable by Bethe in
the 30's.\cite{bethe} This work did not only initiate the 
still thriving field of integrable model systems, but 
founded
the theory of
1D quantum antiferromagnets long before their
experimental discovery in systems such as Sr$_2$CuO$_3$.\cite{ami}
Another example is the Shastry-Sutherland model,\cite{shastry}
whose study predated the discovery of
a closely related valence bond solid ground state
in the compound
 SrCu$_2$(BO$_3$)$_2$.\cite{srcubo1,srcubo2}
Despite these successes, the study of
generic Hamiltonians describing
locally interacting spins on a lattice remains a highly challenging
task. 
The low 
energy properties of a given model are often difficult to extract 
with great confidence. However, the common scenario in dimensions greater than 
one is that the ground state of a 
system of lattice spins governed by a local Hamiltonian will
display some form of long-range order, by either breaking $SU(2)$-spin
rotational symmetry or lattice translational symmetry, or both
(see, e.g., Ref. \onlinecite{RS}).
This phenomenon is well understood through the
general framework that has been
developed around Landau's notion of spontaneous symmetry breaking.\cite{landau}
In recent decades, however, workers in the field have increasingly been
interested in conditions that allow the ground state of a spin system
to remain quantum disordered.
Anderson has argued that a quantum magnet may refrain from 
symmetry breaking even at $T=0$
as a result of quantum fluctuations and/or
frustration.\cite{anderson1}
He later proposed that the resulting ``resonating
valence bond spin liquid'' state may be thought
of as the ideal parent state for the cuprate superconductors.\cite{anderson2}
This proposal has lead to considerable efforts 
in searching for
such a state. Both on the theoretical and on the
experimental side this search turned out to be
a formidable challenge.
Experimentally, there has been much recent excitement about
compounds featuring two dimensional layers
of spin-$\frac 12$ degrees of freedom forming
a triangular\cite{triangular1, triangular2} or kagom\ea\cite{kagome1,kagome2, kagome3,kagome4}
lattice, with no apparent sign of order at low temperatures.
On the theoretical side, a standard way to 
establish the existence of a phase is to
identify a low energy effective field theory
describing the universal properties
of the phase,  
together with special solvable points 
in the phase diagram of some microscopic
Hamiltonian that can be demonstrated 
to display some of these universal features.
This strategy has been very successful 
in a variety of contexts, such as interacting 1D
quantum systems or the fractional quantum Hall effect.
While a thorough understanding of the phase 
is usually 
possible only through the field theoretic
description, its {\em existence} in a certain
microscopic setting may be questionable
until a microscopic realization is found,
either in theory or in experiment.
This is due to the fact that mappings
between microscopic and field theoretic
descriptions, while extremely powerful,
are necessarily non-rigorous. On the other hand,
the construction of exactly solvable higher dimensional
spin-Hamiltonians is generally difficult.
Recent successes along these lines with regard
to the spin liquid problem will be reviewed in
the next subsection.
Currently, however, there is (to the best of my knowledge)
no concrete example for an $SU(2)$-invariant spin-$\frac 12$ Hamiltonian
on a simple lattice,
whose ground state properties are analytically accessible
and agree with those of a gapped (topological) spin liquid.
The main purpose of this paper is to propose such a
Hamiltonian. 

%especially with regard to the problem of ($SU(2)$-invariant) spin liquids .
%Exceptions, both with and without $SU(2)$-invariance, will be discussed in the 
%following subsection.

\subsubsection{Klein models and quantum dimer models} 

The difficulty of constructing solvable models
of spin liquids motivated Kivelson and Rokhsar
to consider the problem in a simplified
Hilbert space.\cite{RK} Their model features
``dimer'' degrees of freedom on the links
of a lattice, which represent singlet bonds between
nearest neighbor sites.
These singlets are the ``valence bonds''
of Anderson's original proposal.\cite{anderson1}
Endowed with simple dynamics, the quantum 
dimer models display phase diagrams that
may be interpreted in terms of singlet 
spin states with short range spin-spin correlations.
(For recent reviews, see Refs.\onlinecite{ML,moessnerraman}).
While the original quantum dimer model
on the square lattice (as well as other bipartite lattices)
does have an exactly
solvable point with a liquid ground state,
this turned out to be a critical point separating
``valence bond solid'' phases with broken
translational symmetry.\cite{Sachdev89, Leung, MSChandra, MSFradkin} 
However, over a decade later,
Moessner and Sondhi found that a similar model on the
triangular lattice has a solvable
point corresponding to a stable quantum liquid phase.\cite{MS}
Subsequently, similar findings were also made for the
kagom\ea lattice.\cite{misguich}
While these findings are realizations of topologically
ordered\cite{wenniu} quantum liquids in lattice models,
their implications for quantum spin systems are
not immediate. This is not only due to the truncation
of the Hilbert space to nearest neighbor valence bond  (NNVB) states,
but perhaps even more so due to the assumed
orthogonality of different dimer coverings in quantum dimer models.
In contrast, the associated valence bond states are not
orthogonal. In fact, even their linear independence is far 
from obvious. Though the linear independence of NNVB states
has so far been
proven only
for the square and honeycomb lattice by Chayes, Chayes,
and Kivelson,\cite{kcc} it is assumed to hold more generally, e.g. based on numerical studies.\cite{mambrini, ML2}
This matter is of fundamental interest in any attempt to formulate
effective theories for frustrated spin-$1/2$ magnets within
the NNVB subspace, except perhaps when a large-N point of view\cite{RS2}
is adopted.
Furthermore, 
arguments assuming linear independence properties 
of NNVB states
have been employed in entropic considerations
for frustrated spin-$1/2$ systems
e.g. on the kagom\ea lattice,\cite{elser_li} 
as well as others.\cite{nussinov_pyro}
For the kagom\ea case, 
the linear independence of the nearest neighbor
valence bond states will be established
in this work.

Although exact mappings between the dynamics of quantum
dimer models
and that of $SU(2)$-invariant spin-$1/2$ quantum magnets on the same
lattice do not exist, 
more general
mappings have been
applied successfully. In Ref. \onlinecite{balents}, a dimer
model
on the triangular lattice has been mapped onto a model of spin degrees
of freedom in a highly anisotropic kagom\ea antiferromagnet.
A mapping that preserves $SU(2)$-invariance has been performed
in Ref.\onlinecite{Sondhi_decorate}, where highly decorated lattices
are considered.
It is interesting to note that for some higher spin systems,
the problem of writing down $SU(2)$-invariant 
Hamiltonians with unique spin liquid ground states
has been solved over 20 years ago through the well known AKLT construction.\cite{AKLT}
These states have been shown to be gapped in 1D,\cite{AKLT2}
and are believed to be gapped in higher dimensions as well.
In contrast, through generalizations\cite{oshikawa, hastings} of the
Lieb-Schultz-Mattis theorem,\cite{LSM} 
it is nowadays well understood
that spin-$1/2$ liquids must be either gapless or have
a non-trivial topological ground state degeneracy 
on lattices with an odd number of sites per unit cell
(cf. also Refs. \onlinecite{affleck, haldane, read, bonesteel}).
The latter implies that they are topologically ordered.\cite{wenniu}
There is much interest in such topologically ordered
phases recently, motivated in part  
by their proposed use in quantum computation.\cite{kitaev1}
A significant number of solvable lattice models with topologically
ordered ground states are known.\cite{kitaev1, kitaev2, freedman2,levin,fendley}
Some of these models can be
naturally cast in terms of spin-$1/2$ degrees of freedom, but will
lack $SU(2)$-invariance in this language.
On the other hand, a parent Hamiltonian
for a topologically ordered $SU(2)$-invariant {\em chiral}
spin liquid state has recently been discussed,\cite{Schroeter, Thomale}
where time reversal symmetry is explicitly broken.

An alternative route to the
construction of $SU(2)$-invariant models with spin liquid ground states
is based on a class of models introduced by Klein.\cite{klein} The Klein-models
have an extensive ground state entropy, with all NNVB states being ground states. When perturbed, one expects
the ground state(s) to be a coherent superposition of NNVB states, to good approximation. This is particularly so if
{\em all} ground states at the Klein point are of valence bond type.\cite{kcc}
It is, however, non-trivial to determine the nature of such perturbed
Klein models. 
%One example where this has been achieved will be discussed
%in the following subsection.

\subsubsection{Outline}

In this work, a local $SU(2)$-invariant 
spin-$\frac 12$ Hamiltonian will be constructed on the
kagom\ea lattice, which has ground states
that are spin-$\frac 12$ realizations of the 
dimer liquid at the solvable ``Rokhsar-Kivelson'' (RK)
point of the quantum dimer model on the same lattice.\cite{misguich}
Some rigorous statements about the uniqueness of these
ground states are deeply related to the linear independence
of the NNVB states
on general kagom\ea type lattices. 
As a byproduct, the linear independence of these
states will be proven in the beginning of this paper. In Section \ref{defs},
this linear independence property will be stated
precisely, and various definitions are introduced
that will be useful in the remainder of the paper.
The linear independence property is then proven
in section \ref{li1}. 
The proof is based on the observation that the linear independence
property for quite general and arbitrarily large kagom\ea-type
lattices can be reduced to a property of finite clusters.
This property can be verified  numerically, or alternatively,
by analytic means.
Both methods have been carried out,
although the details of the analytic method 
(for which there is no real need except from a purist viewpoint)
are not 
presented here to keep the length of the paper within
certain bounds.
Possible generalizations of the linear independence theorem 
and application of the present method to different lattice types
are then discussed in Section \ref{gen1}.
Section \ref{Hamilt} is devoted to the construction of local
$SU(2)$-invariant Hamiltonians with valence bond-type
ground states. In Section \ref{klein}, Hamiltonians
are constructed such that any NNVB state on the kagom\ea lattice is a
ground state, and, unlike for the Klein model
on the same lattice,
there are no {\em obvious} other ground states.
In sections \ref{RVB}1-3, this construction is further generalized
to allow no obvious ground states other than the Sutherland-Rokhsar-Kivelson (SRK)
wavefunctions which are akin to similar states
at the exactly solvable point of the kagom\ea lattice
quantum dimer model.\cite{misguich} In Section \ref{uproof} it is then proven
that these are in fact the only ground states within the
manifold of NNVB states.
In Section \ref{gen2},  a strategy will be discussed
to prove that the SRK wavefunctions are the
only ground states even within the full Hilbert space.
This question will be related to properties of the
generalized Klein models of Section \ref{klein}, which are
not proven here. In Section \ref{discussion} it will be argued
that the model constructed in Section \ref{RVB} likely
describes a $\mathbb{Z}_2$ topological liquid.
%This conjecture is based on the assumption that
%properties of the correlation functions carry over from
%the related quantum dimer model, as discussed
%below and in the following, and on the uniqueness
%of the four topologically degenerate ground states,
%which is only partially proven here (i.e. within the
%valence bond subsector). 
Section \ref{conclusion}
concludes the paper. A technical detail regarding
the uniqueness of the SRK ground states is
presented in the Appendix.

I finally remark upon the relation of the present work to
that of Ref. \onlinecite{fujimoto},
where
spin-Hamiltonians whose ground states
are valence bond realizations of 
RK dimer states
have also been constructed.
In Ref. \onlinecite{fujimoto}, the corresponding dimer
states describe critical points in the phase
diagram of their dimer Hamiltonians.
In contrast, the dimer wavefunction that inspired 
the ground states of the model constructed here
corresponds to a stable $\mathbb{Z}_2$ topological
liquid. 
In section \ref{discussion}, it will be argued
that the properties of the dimer liquid will largely
carry over to the spin wavefunction. Similar arguments
have been made for the critical cases in Ref. \onlinecite{fujimoto},
and have been put forth early on by Sutherland\cite{sutherland}
for the square lattice case.
The strategy developed here is applicable to RK-points
on different lattices as well, including the square and honeycomb cases of Ref. \onlinecite{fujimoto}, although the
present construction would not in any obvious way
give rise to the same Hamiltonians discussed there.
Application of the present construction  to other lattices
will be further discussed in
Section \ref{gen2}.

\section{Definitions and statement of the linear independence property\label{defs}}
\begin{figure}[b]
\includegraphics[width=7cm]{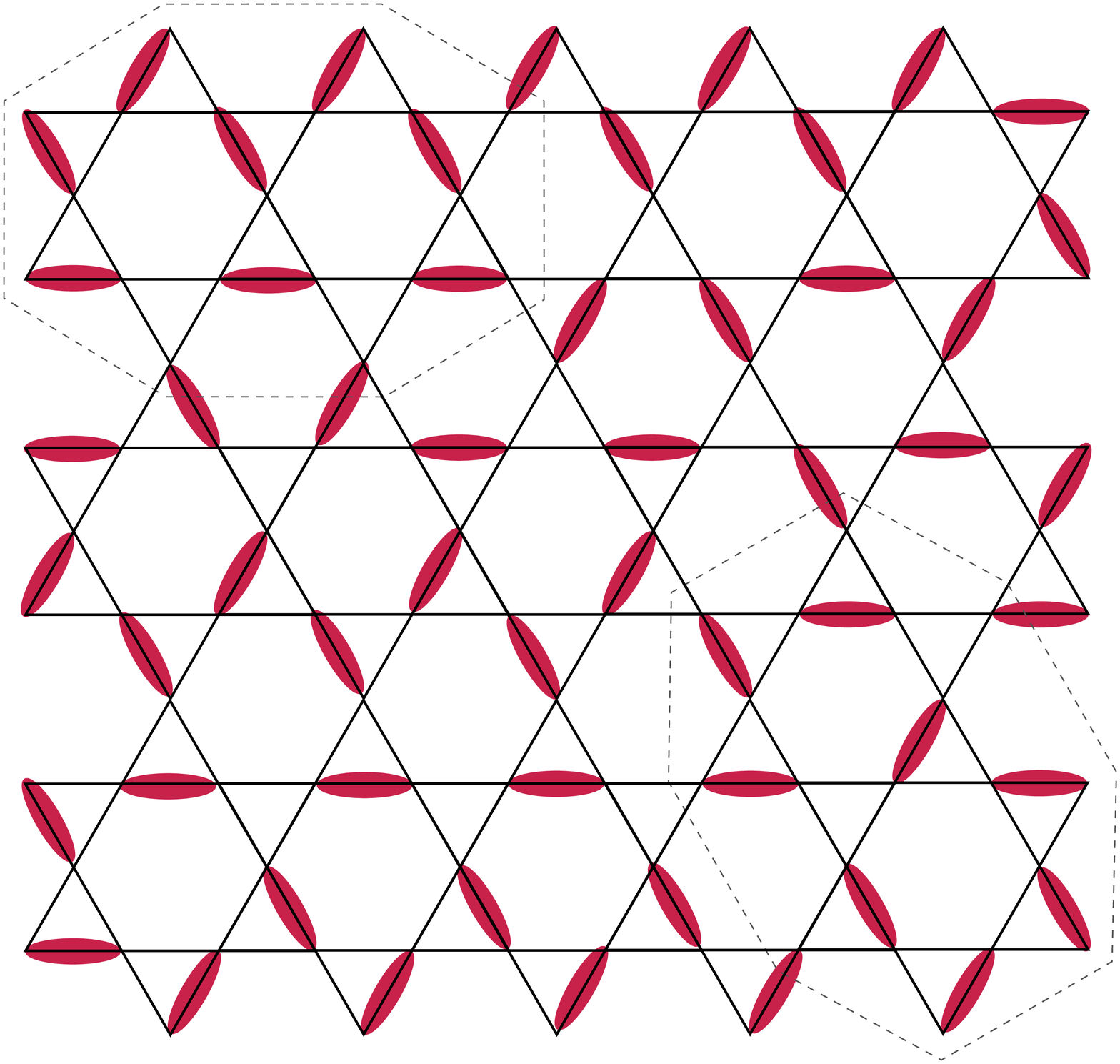}
\caption{\label{kagome} A kagom\ea lattice. Some 19-site cells with the
topology of Fig. (\ref{cells}b) are indicated. The lattice is ``regular'' as defined in the text. It is also fully dimerizable, as indicated by the shown dimer covering.}
\end{figure}
I will start by defining some terms that will be useful in
the following.
In general, I will use the term {\em lattice} to refer to any collection
of discrete points or lattice sites, $\cL$, which have
a topology imposed through the notion of {\em nearest neighbors}.
The set of nearest neighbors of a lattice site $i$ is called
its {\em neighborhood}, and is denoted by $\cN(i)$. If $j$ is a
nearest neighbor (NN) of $i$, then $i$ is a NN of $j$, and
the (unordered) pair $(i,j)$ is called a link of the lattice. 
I will say that $i$ and $j$ are the sites ``touched'' by
the link $(i,j)$.
While many aspects of this work are expected to generalize
to other lattices of interest, I will mainly focus on kagom\ea-type
lattices  (\Fig{kagome}). For now, the lattice $\cL$
may be any finite subset of the infinite two-dimensional (2D)
kagom\ea lattice, or, more generally, of an arbitrarily
large but finite kagom\ea lattice with toroidal periodicity.  
More general
global topologies are also possible as long as
the local structure is that of a kagom\ea lattice,
as will become clear in the following. For greater
simplicity, however, I will defer the discussion
of such cases to Section \ref{gen1}.

\begin{figure}[t]
\centering
\includegraphics[width=8cm]{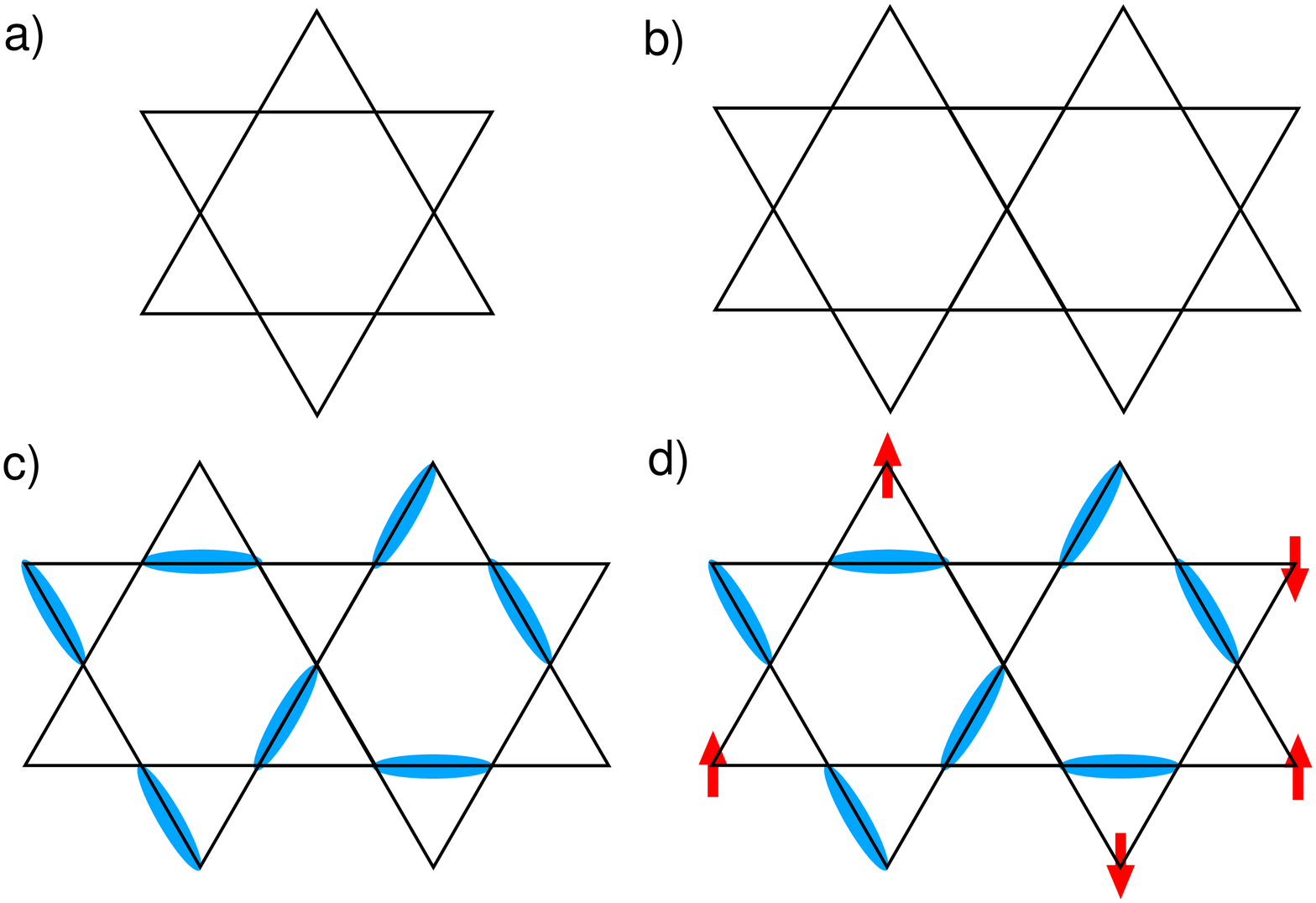}
\caption{\label{cells} a),b) Some cells of the kagom\ea lattice. 
The cell shown in b) is the smallest cell for which the linear independence
property stated in the text holds. It is thus the 
smallest
``regular'' kagom\ea lattice as defined in the text.
c) A dimer covering $D$ of the cell shown in b). Note that according
to the definition used here, boundary sites need not participate.
d) A state 
compatible 
with the dimer covering $D$ shown in c), i.e.,
an element of
$\cH(\ket{D})$. Dimers correspond to singlets formed by
spin-$1/2$ degrees of freedom on neighboring sites, whereas
boundary sites not touched by dimers are allowed to be in
an arbitrary state. }
\end{figure}

A {\em cell} may be any subset $\cC$ of a given lattice $\cL$,
but will usually refer to reasonably small and well connected
units such as shown in \Fig{cells}, 
with non-vanishing interior (to be defined next). 
For kagom\ea-type 
lattices, I define the {\em interior} $\o{\cC}$ of the cell $\cC$ as
the set of all points in $\cC$ that have four nearest neighbors 
also contained in $\cC$. Likewise, the {\em boundary} $\partial \cC$ of 
 $\cC$ consists of those points in $\cC$ that are not interior.
A {\em dimer covering} $D$ of the cell $\cC$ is a set of
{ disjoint (!)}
links between sites in $\cC$
such that 
each interior site of $\cC$ is touched by one link in $D$.
I will also refer to the links of $D$ as the {\em dimers}
of the covering.
The support of the dimer covering $D$, denoted $\supp(D)$,  is the
set of all sites belonging to a dimer in $D$.
Note that $\o{\cC}\subset\supp(D)$ by definition, but the
opposite inclusion need not hold, since $\supp(D)$
will in general contain some boundary points of $\cC$ as well, cf. \Fig{cells}. 
Likewise, 
I introduce the set of {\em free sites} $\free(D,\cC)=\cC\backslash\supp(D)$ of the dimer covering $D$, which contains all (boundary)
sites in $\cC$ that are not touched upon by the covering. The dependence on $\cC$ may be suppressed when it is clear what cell is referred to, and I then denote the free sites of the covering simply by $\free(D)$.
By $\cD(\cC)$ I denote the set of all dimer
coverings of $\cC$, and by $\cD\equiv\cD(\cL)$ the set of all
dimer coverings of the lattice.
For every subset $\cS$ of $\cL$, $\cH(\cS)$ denotes the Hilbert space
obtained by associating a spin-$1/2$ degree of freedom 
with every site in $\cS$, and $\cH\equiv\cH(\cL)$ the Hilbert
space associated with the entire lattice. Since $\cH(\cS)\cong\mathbb{C}^{2^n}$,
in general, where $n$ is the number of sites in $\cS$, it is natural 
to take $\cH(\cS)\cong\mathbb{C}$ should $\cS$ by some chance be empty.
With this convention, we can always write 
\begin{equation}
\cH(\cC)=\cH(\supp(D))\otimes\cH(\free(D))
\end{equation}
when $D$ is a dimer covering of $\cC$.
%which holds even if there are no free sites. 
More generally, consider a cell $\cC$ that is the disjoint union of two subcells $\cC'$ and $\cC''$, such that $\cH(\cC)=\cH(\cC')\otimes\cH(\cC'')$. Given a (pure) state $|s'\rangle\in\cH(\cC')$, we can define a subspace $\cH(\ket{s'},\cC)$ of $\cH(\cC)$ 
consisting of all states that are {\em compatible} with the state $\ket{s'}$:
\begin{equation}
\begin{split}
  \label{comp}
  \cH(\ket{s'},\cC)  &=\text{span}(\ket{s'})\,\otimes\,\cH(\cC'')\\
 &=\{\ket{s'}\otimes\ket{s''}:\ket{s''}\in \cH(\cC'')\}
  \,,
\end{split}
\end{equation}
where span$(\cdot)$ denotes the linear span of the vector(s)
enclosed by the brackets.
Again, if it is clear what cell is referred to, we may write
$\cH(\ket{s})$ instead of $\cH(\ket{s},\cC)$. See \Fig{cells}
for an example.

Let us fix a cell $\cC$ for the moment. 
For a given dimer covering $D$ of $\cC$ we denote by 
$|D\rangle$ a state 
in $\cH(\supp(D))$ in which any two spins belonging to the same link in $D$ form a singlet
or ``valence bond'' (VB).
The state $|D\rangle$ is just a realization of the dimer covering $D$
through valence bonds. Then $\cH(\ket{D})$, as defined in \Eq{comp},
consists of all states that are compatible with the dimer covering $D$. A general state belonging to $\cH(\ket{D})$ for some
dimer covering $D$ will have every internal site participating in a nearest neighbor singlet, while boundary sites left ``free'' by the covering $D$ may be in any arbitrary state (cf. Fig. (\ref{cells}d), but note that the free sites could in general be entangled). The space spanned by all states of this kind will be called the space of {\em valence bond states} of the cell $\cC$, denoted by $VB(\cC)$:
\begin{equation}
  \label{VB}
  VB(\cC)=\sum_{D\in \cD(\cC)} \cH(\ket{D})\;,
\end{equation}
where the sum denotes the linear span of the spaces summed over.
Note that the space $VB(\cC)$ is $SU(2)$-invariant. This is so since
for $\ket{s'}=\ket{D}$, both factors in the first line of \Eq{comp} are 
$SU(2)$-invariant, and hence $\cH(\ket{D})$ is $SU(2)$-invariant for each $D$.

It is clear from the definition how to write down a set of states that linearly generate the space $VB(\cC)$.
For each dimer covering $D$, we denote by 
$|\psi_{D,j}\rangle$ a 
basis of the space $\cH(\free(D))$ of spins not touched by the covering. Here, $j$ runs from 1 to $n_D\equiv2^\wedge|\free(D)|$.
By definition, then, the following set of states,
\begin{equation}
  \label{B}
  \cB(\cC)=\{\ket{D}\otimes\ket{\psi_{D,j}}:D\in\cD(\cC), j=1\dotsc n_D\}\,,
\end{equation}
linearly generates the space $VB(\cC)$ of valence bond states on the
cell $\cC$.
 We will be interested in the question for what cells $\cC$ all states in $\cB(\cC)$ are linearly independent.
Note that if $\cC$ is the entire lattice $\cL$, and $\cL$ has no boundary, then there are no free sites for any $D$, and the valence bond states
$\cB(\cL)$ consist of all NN valence bond ``dimerizations'' of the lattice
in the usual sense. 
In the presence of a non-trivial boundary, however, states are also admitted where some boundary sites do not participate in the NN singlet bonding. Since more states are being included, the statement of their linear independence is a stronger one than if only ``true'' dimerizations were considered. 
Indeed, the great advantage of including these states
is that the linear independence of the set $\cB(\cC)$ for certain small cells $\cC$ is now so strong 
a statement 
that it immediately carries over to entire 
lattice, as will be demonstrated shortly below. 
This fact, and the observation that some fairly small cells already have this linear independence property, constitute the main ingredients of the proof 
constructed in the following Section.
\begin{figure}[t]
\centering
\includegraphics[width=8cm]{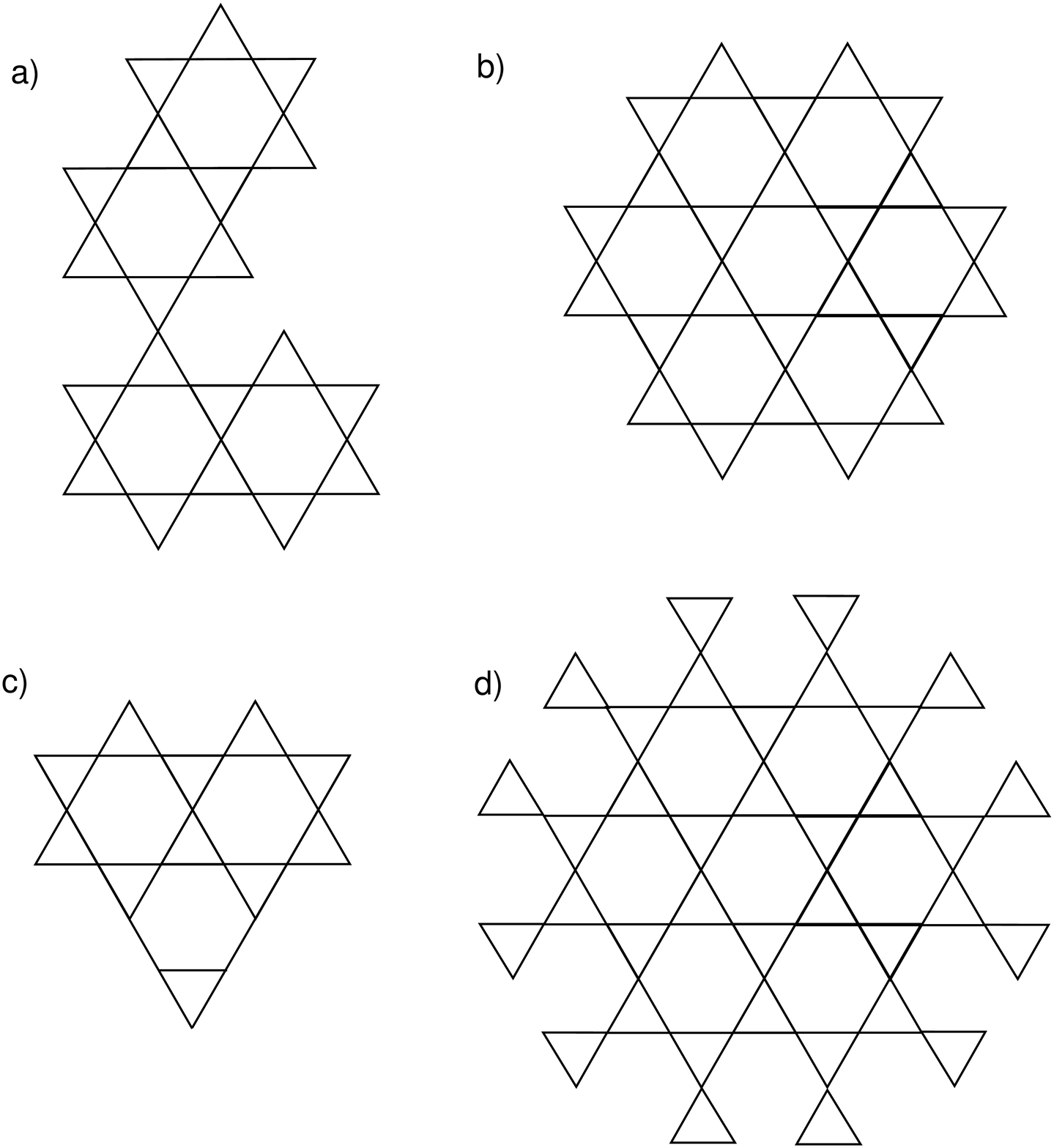}
\caption{ 
Regular and non-regular kagom\ea lattices. 
a)+b) Regular kagom\ea lattices. b) is fully dimerizable,
a) is not, due to an odd number of sites.
c)+d) Non-regular kagom\ea lattices.\label{regular}
 }
\end{figure}
For simplicity, I will now focus on lattices $\cL$ that are subsets of some finite 2D kagom\ea lattice with doubly periodic (toroidal) boundary conditions. Any such lattice will be called a {\em regular kagom\ea lattice}, if and only if any link $(i,j)$ of $\cL$ 
 belongs to a 19-site cell $\cC$ contained in $\cL$ that has the topology shown in Fig. (\ref{cells}b).
%any {\em inner} site $i$ of $\cL$ 
%is also an {\em inner} site of some such cell.
In general, the cell $\cC$ will of course depend on the link $(i,j)$
and need not be unique.
Examples for regular kagom\ea lattices
include such interesting cases as
the lattice depicted in \Fig{kagome}, those in Figs. (\ref{regular}a,b), as well as any
sufficiently large kagom\ea lattice with toroidal periodic
boundary conditions.
The following theorem is the main result on linear 
independence of valence bond states, and will be proven in
Section \ref{li}:

{\em Theorem I:} For any regular kagom\ea lattice $\cL$, the set of valence bond states $\cB(\cL)$ is linearly independent.

Furthermore, one is sometimes
interested only in {\em full dimerizations} of the
lattice, where every site is touched by a dimer, including
all boundary sites. I will call the lattice {\em fully dimerizable}
if and only if there is a way to group all sites into disjoint pairs
of nearest neighbors. Any such way will be called a full dimerization.
Full dimerizations of regular lattices are special cases
of dimer coverings as defined above.
The corresponding valence bond states
are thus a subset of the set $\cB(\cL)$, and one obtains the following theorem
 as simple corollary of Theorem I:

{\em Theorem II:} For any fully dimerizable regular kagom\ea
lattice, the valence bond states associated with full dimerizations
are linearly independent.

\section{Proof of the linear independence property\label{li}}

\subsection{Regular kagom\ea lattices\label{li1}}

One immediate consequence of Theorem I is that the 19-site cell shown in 
Fig. (\ref{cells}b) does by itself already have the stated linear independence property. 
In this Section, $\cC$ will always refer to a cell of this topology.
Indeed, this seems to be the smallest cell of the kagom\ea lattices for which the 
set $\cB(\cC)$, \Eq{B}, is linearly independent
(see the Appendix for ``spoilers'' when $\cC$ is chosen to be the cell
in Fig. (\ref{cells}a)). 
%The proof that the 19-site cell has this linear independence property will be deferred to the Appendices. 
Conversely, it will be shown in the following that once this property is established for the 19-site cell, it immediately generalizes to any kagom\ea lattice that is regular in the sense defined above. 
Although I will focus on the kagom\ea case, the argument for this
is very general, and should allow for rather direct generalizations to other cases of interest: Once the linear independence of $\cB(\cC)$ is established for suitable ``building blocks'' $\cC$ of the lattice, it will generalize to the lattice as a whole. 
I will call these building blocks the ``bricks of linear independence''
of the lattice $\cL$. 
These bricks may quite generally be reasonably small, as is at least suggested by the present example: The set $\cB(\cC)$ for the 19-site cell in Fig. (\ref{cells}b) consists of 13120 states, whose linear independence can be verified numerically or using computer algebra in a straightforward manner. 
This task amounts to checking that a suitably defined integer overlap matrix has 
full rank. I have used the LinBox package\cite{LinBox} for this purpose.
However, an analytic proof of the linear independence of $\cB(\cC)$ is also possible,\cite{ASU} using the results stated in the Appendix
for the 12-site cell of Fig. (\ref{cells}a). 
The latter can be derived
using Rumer-Pauling valence bond diagrams.\cite{rumer,pauling,soos1,soos2}
It is of little importance, however, whether the linear
independence property of the set $\cB(\cC)$ is obtained
by analytic or numerical means. Once this property
is established, the same property follows for arbitrarily
large regular lattices, as will be shown in the following.

According to the above, the set $\cB(\cC)$ defines a basis for the space of valence bond states $VB(\cC)$, \Eq{VB}. If we label these basis states 
by an index $b$ and write $\ket{b}$ for the elements of $\cB(\cC)$, we can therefore
define a set of linear projection operators
$P_b$ with the following properties:
\begin{equation}\label{Pb}
  P_b\ket{b'}=\delta_{b,b'}\ket{b}\,,\quad
  P_b P_{b'}= \delta_{b,b'} P_b\;.
\end{equation}
The action of the operators $P_b$ within the space $VB(\cC)$
is fully defined by \Eq{Pb}.\cite{note1}
The existence of such operators is guaranteed by the linear
independence of the set $\cB(\cC)$.
Note that the $P_b$ are not
Hermitian, since the states $\ket{b}$ are not orthogonal.

As a next step, I define projection operators that leave
all valence bond states corresponding to a given dimer
covering $D$ invariant, and annihilate all valence
bond states corresponding to dimer coverings $D'\neq D$.
This is easily accomplished in terms of the operators
defined by \Eq{Pb}. Recall that each of the states
$\ket{b}\in \cB(\cC)$ is of the form $\ket{b}=\ket{D}\otimes\ket{\psi_{D,j}}$,
where $\ket{\psi_{D,j}}$ denotes the state of the sites not touched
by the dimer covering. We can thus write $P_b\equiv P_{D,j}$.
The operators
\begin{equation}\label{PD0}
  P_D=\sum_j P_{D,j}
\end{equation}
then satisfy
\begin{equation}\label{PD}
  \begin{split}
     P_D\;\ket{D'}\otimes\ket{\psi_{D',j}}&=\delta_{D,D'}\,\ket{D'}\otimes\ket{\psi_{D',j}}\,,\text{for any }j,\quad\\
     P_D P_{D'}&= \delta_{D,D'} P_D\;.
  \end{split}
\end{equation}
The idea is now to observe that the successive action
of operators of this kind defined on various 
19-site bricks
of some larger regular lattice $\cL$ can ``single
out'' any given dimer covering of this lattice.
Detailed arguments are given in the following.

Consider now a regular kagom\ea lattice $\cL$. 
%I will now refer to any 19-site cell $\cC$ of 
%$\cL$ that has the topology of Fig. (\ref{cells}b)) a ``brick''
%of the lattice. 
By definition, every link of the regular lattice
belongs to at least one 19-site brick of the topology
shown in  Fig. (\ref{cells}b). 
For definiteness, one may consider the
brick shown in the upper left hand corner of the lattice in \Fig{kagome}.
Let now $D_\cC$, $D'_\cC\in \cD(\cC)$ be dimer coverings of the brick $\cC$. 
Further consider any state on $\cL$ that is compatible with the dimer covering
$D_\cC$ in the sense that it factorizes into the valence bond state
$\ket{D_\cC}$ on $\supp(D_\cC)$ and any other state $\ket{S}$
on $\cL\backslash\supp(D_\cC)$, i.e. a state of the form
$\ket{D_\cC}\otimes\ket{S}$. The properties \Eq{PD} of the operators
$P_{D'_{\cC}}$, which act on the brick $\cC$, immediately imply
\begin{equation}\label{PD2}
 P_{D'_\cC} \, \ket{D_\cC}\otimes \ket{S} = \delta_{D_\cC,D'_\cC} \,\ket{D_\cC}\otimes \ket{S}\,.
\end{equation}
To see this, all we need to do is to expand the state
$\ket{S}$ in a basis of the form $\ket{\psi_{D_\cC,j}}\otimes\ket{S'_{j'}}$,
where the first factor is a basis state of $\cH(\free(D_\cC,\cC))$
as it appears in \Eq{PD}, and the second factor is an element
of some basis of $\cH(\cL\backslash\cC)$. \Eq{PD2} then immediately
follows from \Eq{PD}.

Next consider a particular dimer covering $D$ of $\cL$,
and a valence bond state compatible with $D$, $\ket{D}\otimes\ket{\psi_{D,j}}$.
Again, $\ket{\psi_{D,j}}$ is a 
state of $\cH(\free(D,\cL))$, chosen from some arbitrary basis.
The dimer covering $D$ induces a dimer covering of $\cC$, 
the {\em restriction}
$D_\cC$ of $D$ on $\cC$, consisting of all dimers of
$D$ that are fully contained in $\cC$ (cf., e.g., \Fig{kagome}).
%The valence bond state $\ket{D}\otimes\ket{\psi_{D,j}}$ clearly
%has the property that 
%the sites in $\supp(D_\cC)$ 
%(i.e. the sites touched by the dimers in $D_\cC$), are not entangled with 
%the rest of the lattice. 
Then, the  valence bond state $\ket{D}\otimes\ket{\psi_{D,j}}$ is of the form
$\ket{D_\cC}\otimes \ket{S}$ displayed in \Eq{PD2}.
To make this explicit, one may introduce 
the {\em complement} of the
dimer covering $D_\cC$ in $D$, which consists of those dimers in $D$ that
are not contained in $D_\cC$, and denote it by $\overline{D_\cC}$.
Then by definition 
\begin{equation}
  \label{tensor}
  \ket{D}=\ket{D_\cC}\otimes\ket{\overline{D_\cC}}\;.
\end{equation}
Hence, the state $\ket{D}\otimes\ket{\psi_{D,j}}$ 
is 
of the stated form with $\ket{S}=\ket{\overline{D_\cC}}\otimes\ket{\psi_{D,j}}$, and 
from \Eq{PD2} we have
\begin{equation}\label{PD3}
   P_{D'_\cC} \, \ket{D}\otimes\ket{\psi_{D,j}} = \delta_{D_\cC,D'_\cC} \,\ket{D}\otimes \ket{\psi_{D,j}}
\end{equation}
for any dimer covering $D'_\cC$ of the brick $\cC$, where again
the operator $P_{D'_\cC}$ acts on this 
particular brick. In the following, the case of interest
will be that $D'_\cC$ is also obtained as the restriction
on $\cC$ of some dimer covering $D'$ of $\cL$.
In words, \Eq{PD3} then says that the valence bond state 
$\ket{D}\otimes\ket{\psi_{D,j}}$ will survive the action of 
$P_{D'_\cC}$ unaltered if the dimer coverings $D$ and $D'$
locally look the same within the brick $\cC$, and will be
annihilated if not. Here, ``looking the same locally'' means
that all dimers fully contained in $\cC$ are identical.

The proof of the linear independence of the set $\cB(\cL)$
is now trivial. Suppose we have a linear combination of
states in $\cB(\cL)$ that vanishes identically:
\begin{equation}\label{lc}
  \sum_{D\in\cD(\cL)}\sum_{j=1}^{n_D}\;\lambda_{D,j}\, \ket{D}\otimes\ket{\psi_{D,j}}=0 \,.
\end{equation}
We need to show that this implies that all $\lambda_{D,j}$ 
are zero. First we successively act on \Eq{lc} with
all operators of the form $P_{D'_\cC}$ for a {\em fixed}
dimer covering $D'$, and for all bricks $\cC$ of the lattice,
i.e., we act on \Eq{lc} with
\begin{equation}\label{proj}
  \prod_\cC P_{D'_\cC}\quad.
\end{equation}
This will eliminate all $D$ with $D\neq D'$ from \Eq{lc}.
For, if $D\neq D'$, there must be a link of the lattice
which, say, belongs to $D$ but not to $D'$. Since the lattice
is regular, there is a brick $\cC$ containing this link.
The operator $P_{D'_\cC}$ corresponding to this brick
will then annihilate all states of the form $\ket{D}\otimes\ket{\psi_{D,j}}$.
Note that the operators $P_{D'_\cC}$ commute within $VB(\cL)$,
since by \Eq{PD3}, $VB(\cL)$ is the span of a common set of eigenstates.
It follows, then, that the action of \Eq{proj} on \Eq{lc}
annihilates all states with $D\neq D'$,
whereas
all states of the form
$\ket{D'}\otimes\ket{\psi_{D',j}}$ are invariant under this action.
We are thus left with an equation of the form
\begin{equation}
  \sum_{j=1}^{n_{D'}}\;\lambda_{D',j}\, \ket{D'}\otimes\ket{\psi_{D',j}}=0 \,.
\end{equation}
But since the states appearing in here only differ through the
states $\ket{\psi_{D',j}}$ on $\free(D')$, and the states $\ket{\psi_{D',j}}$
are linearly independent by definition, it follows that $\lambda_{D',j}=0$
for any $j$. Since $D'$ was arbitrary, all coefficients in 
\Eq{lc} must vanish identically. 

This completes the proof 
of Theorem I.

%hold for arbitrary regular lattices if it holds for the 19-site 
%cell in Fig. (\ref{cells}b). 

%The proof for the 19-site cell
%is given in the Appendices. It is more laborious but
%can be replaced easily by a numerical effort, as mentioned above.

\subsection{Further generalizations\label{gen1}}

For reasons of simplicity, Theorems I and II have not been stated
in the most 
general form one could imagine. 
For one, the restriction to sublattices of finite periodic 
kagom\ea lattices, which have toroidal topology, is not
strictly necessary. This was done because such lattices
do not require a very technical definition, and are certainly
general enough for most purposes. Quite similarly,
when talking about two-dimensional manifolds one will
think of these as being embedded into three-dimensional
Euclidean space in simple enough cases, though this is not
possible in general. Likewise, if one wants
to apply the present results to a kagom\ea-type
lattice, say, of M\"obius-strip topology, one
would desire a more "local" definition of a regular
kagom\ea lattice. There is, in fact, no real
obstacle in applying the present results to such 
a lattice, since the operators $P_{D_\cC}$
are entirely local.
The only real requirement that must be made
is thus that the lattice is "built up"
from the 19-site bricks defined above.
One must define carefully, however, what "built up"
means with regard to the neighborhood structure.
For example, it is clear that one cannot apply the present results 
to three dimensional stackings of planar kagom\ea
lattices without further ado, if there are links along
the third direction. A sufficiently strong requirement
would be
that the lattice can be covered by bricks,
such that for each lattice site, there is a brick containing
the entire neighborhood of this site.
More generally, it is only necessary that
the restrictions of any dimer covering
to all the bricks of the lattice uniquely
determine the covering.
For kagom\ea type lattices, using the
arrow representation\cite{elser_arrow} (see Section \ref{uproof})
one can show that this is already the case
when each {\em site} of the lattice, rather than
each link, is contained in a brick.
This will be of some importance in Section \ref{uproof}.
(Cf. \Fig{unique} and caption.)
However, instead of trying to state Theorems
I and II for the most general class of lattices,
it may be more efficient to consider "non-regular"
lattices on a case by case basis, and determine
if the construction of the preceding Section can be used
to derive a linear independence property.
For example, the lattice in Fig. \ref{regular}d) is 
certainly not regular, but Theorem II still applies.
That is so because every full dimerization of this
lattice necessarily has dimers on any of the links
between two boundary sites. These dimers are thus
mere spectators, and have no bearing on the question
of linear independence. The remaining lattice is,
however, regular, and so the linear independence of
the valence bond states corresponding to full dimerizations
still holds.

As stated, the present strategy to prove the linear
independence can be readily applied to other lattices,\cite{note5}
provided that elementary ``bricks'' of the lattice can
be identified which cover the lattice in the sense
discussed above, and for which the linear independence
of the set $\cB$ holds. It is in general possible
to have more than one type of brick, which may be 
an advantage if the lattice is somewhat irregular.
Obvious candidates to apply this method to include
the triangular and the "pentagonal"\cite{Sondhi_decorate}
lattice, and also
the square
and the honeycomb lattice in the presence of periodic boundary
conditions, which have so far been studied for open boundary conditions
only. \cite{kcc}
One may be hopeful that the present strategy works at least for lattices
with comparable or lower coordination number compared to the
present case.
While the coordination number of the triangular lattice
is notoriously high, there is no a priori reason
to exclude such cases from consideration. The same
is true for some higher dimensional lattices.
A detailed case by case analysis is left for future studies.

%More general topologies. E.g. ``sphere'', mobius.
%Non-regular lattices, such as \Fig{proper} d).
%Other lattices, e.g. triangular.

\section{A Hamiltonian with Sutherland-Rokhsar-Kivelson-RVB ground states\label{Hamilt}}

\subsection{Generalized Klein Models\label{klein}}

The final goal of this section is to construct a Hamiltonian
whose ground states are special superpositions
of NN valence bond states. One natural starting point
for this endeavor is the construction of a Hamiltonian whose ground
state sector contains the entire manifold of
NN valence bond states, but -- if possible -- no other states.

\begin{figure}[t]
\centering
\includegraphics[width=6cm]{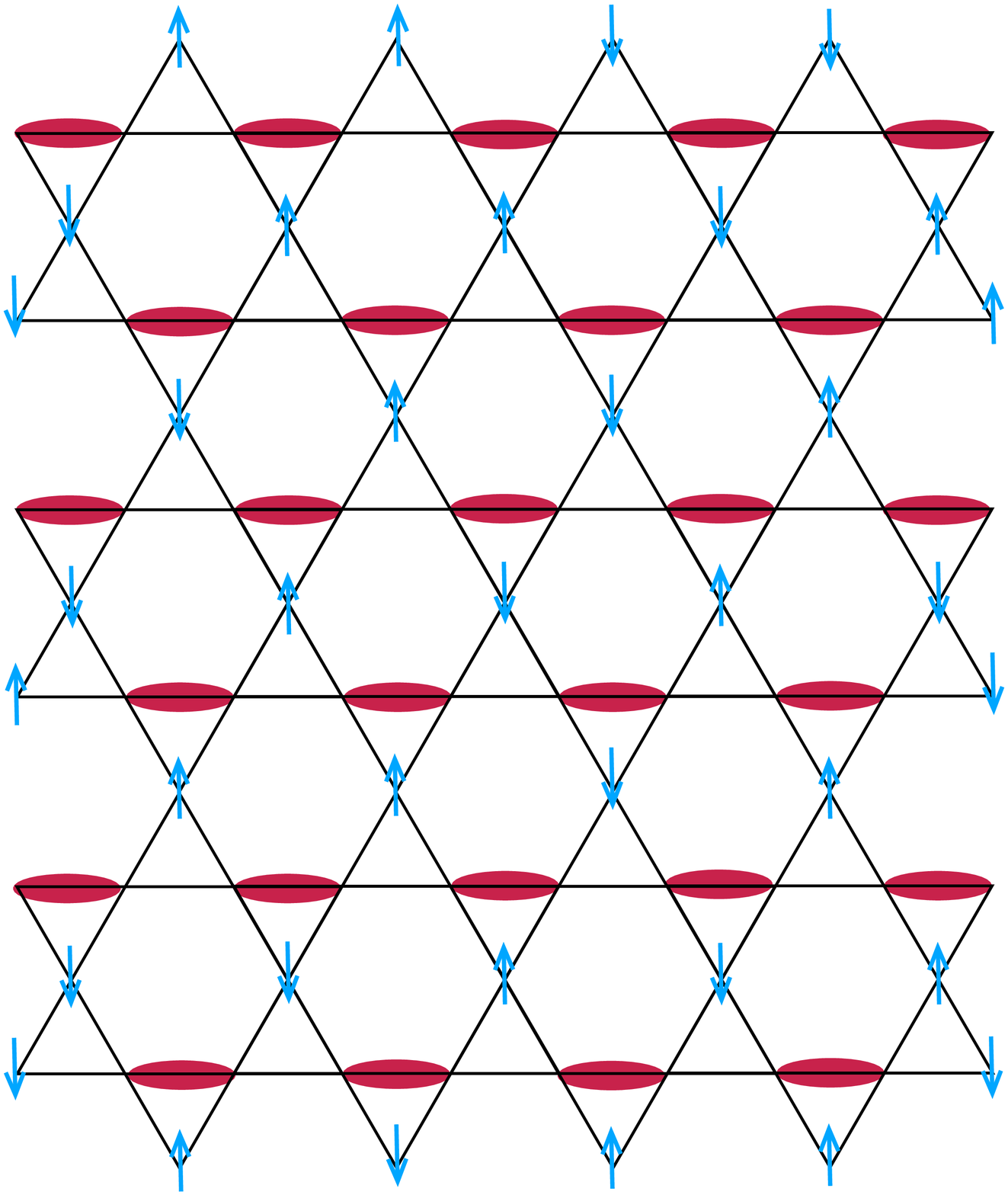}
\caption{ \label{kleings}
A ground state of the Klein model, \Eq{Hkl}, on the kagom\ea
lattice. The state has only one valence bond 
per unit cell, and each bow-tie of the lattice
fully contains one such bond.
The state obviously lies outside the
space of valence bond states, since there 
are many sites not participating in valence bonds.
 }
\end{figure}

In a very influential work\cite{klein}, Klein discussed a 
strategy to construct models of this kind on general lattices.
For each site $i$ of the lattice, a projection
operator $P_i$ is considered which acts on a cell
consisting of the site $i$ and all its nearest neighbors,
and projects onto the subspace of this cell that
has maximum total spin. The Klein Hamiltonian,
\begin{equation}\label{Hkl}
H_{\text{Klein}}=\sum_i P_i\;,
\end{equation}
then has a ground state sector which includes all
NN valence bond states of the lattice. 
In some cases of interest,
including the square lattice and the honeycomb lattice,
it is believed that these valence bond states are a complete
set of ground states of \Eq{Hkl}, and a rigorous proof exists for
the honeycomb case.\cite{kcc} In other cases, however,
there are obvious ground states outside the valence bond
subspace. The kagom\ea lattice is an example of the
latter kind, as demonstrated by the state shown in \Fig{kleings}. 
\begin{figure}[t]
\centering
\includegraphics[width=3cm]{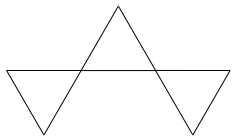}
\caption{ \label{crown}
A 7-site sell. 
The generalized 
Klein Hamiltonian \Eq{gklein} for this cell
no longer has the ground state depicted in \Fig{kleings}.
Some non-valence-bond ground states
do however remain.
 }
\end{figure}
One may ask whether it is possible to construct a local
Hamiltonian for the kagom\ea lattice such that the 
NNVB states
span the entire ground state sector.
Since the Klein Hamiltonian \Eq{Hkl} does not have this property,
one will need to consider projection operators acting
on cells larger than the 5-site ``bow-ties'' formed
by a site $i$ and its nearest neighbors.
Possible examples include the cells shown in \Fig{crown},
and those in Figs. (\ref{cells}a,b).
For larger cells, with more than one internal site,
it is not sufficient to project onto the maximum spin
sector. One solution might be to project onto the
sector with total spin $> N_s/2-N_b$, where 
$N_s$ is the number of sites of the cell, and $N_b$ the
minimum number of valence bonds that must be entirely
contained inside the cell in an arbitrary valence bond states.
Similar constructions, with different motivations in mind,
have been explored in Refs. \onlinecite{Sondhi_decorate, batista, nussinov_pyro, shtengel}.
However, since internal sites may or may not bond with
each other, the number of valence bonds entirely 
contained within a cell may be larger for some valence
bond states than for others. 
In the present context, there is thus a 
more natural, and far more restrictive (for larger cells)
way to construct local projection operators
for which valence bond states are (say) maximum eigenvalue eigenstates.
For any cell $\cC$ consider the {orthogonal} projection operator
$P_{\cC}$ onto the subspace $VB(\cC)$, \Eq{VB}, and the
Hamiltonian
\begin{equation}
  \label{gklein}
H=-\sum_\cC P_\cC\;.
\end{equation}
Here, the sum goes over all cells of a kagom\ea lattice $\cL$
that have a certain topology, e.g. \Fig{crown} or Fig. (\ref{cells}a).
Since $VB(\cC)$ is $SU(2)$-invariant, so is each projection 
operator $P_\cC$ and
thus the Hamiltonian \Eq{gklein}. Furthermore,
for any fixed cell $\cC$, any valence bond state will be a ground
state of the operator $-P_\cC$. 
This is so since any state in $\cB(\cL)$
is of the form $\ket{D_\cC}\otimes\ket{S}$,
with $D_\cC$ a dimer covering of $\cC$ and $\ket{S}$
a state on $\cL\backslash\supp(D_\cC)$,
as explained above \Eq{tensor}. It thus satisfies
\begin{equation}
P_\cC\,\ket{D_\cC}\otimes\ket{S}=\ket{D_\cC}\otimes\ket{S}\;,
\end{equation}
for reasons entirely analogous to those stated below \Eq{PD2}.
Valence bond states are thus ground states of \Eq{gklein}.
If the cells on which the operators $P_\cC$ are defined are just
the 5-site bow-ties, \Eq{gklein} differs from the original Klein Hamiltonian
only by a constant. 
For a sufficiently large choice of cell, however,  
there will be no obvious
non-valence-bond type ground states. 
This motivates the consideration of
\Eq{gklein} for more general cells.
These cells should have the property
that they cover the considered lattice
$\cL$ entirely. Moreover, it is possible
to extend the sum in \Eq{gklein} over 
more than one type of cell.
Furthermore, recall that the definition of valence bond states in Section \ref{defs}
admits states with dangling spins at the boundary $\partial \cL$ of $\cL$, as
long as $\partial\cL$ is not empty.
However, a modification of the boundary terms in
\Eq{gklein} may 
restrict the ground state sector of \Eq{gklein} to
valence bond states corresponding to full dimerizations
of the lattice. Specifically, for cells $\cC$ 
overlapping the lattice boundary $\partial \cL$,
one may restrict the dimer
coverings $D$ of $\cC$ defining the space $VB(\cC)$, \Eq{VB},
to those where every boundary site of $\cL$ is touched
by a dimer. With this modification, only valence bond states
corresponding to full dimerizations will
be obvious ground states of \Eq{gklein}, even if the boundary is not empty.
For a sufficiently large size of the cell $\cC$, it seems likely
that the
valence bond states and their linear combinations will be the unique 
grounds states of \Eq{gklein}.
Note that the linear independence, or lack thereof, of the set
$\cB(\cC)$ is irrelevant for the Hamiltonian \Eq{gklein}
to be meaningful, and is not in any obvious way related to
the question of the uniqueness of the valence bond ground states.
For this reason 
we need not limit our attention to cells
that contain 
the 19-site ``bricks''
studied in the preceding Sections, 
but smaller cells may suffice. 
For example, when the 7-site cell depicted in
\Fig{crown} is used to define the Hamiltonian
\Eq{gklein}, some non-valence-bond
grounds states of the original Klein model
will cease to be ground states of \Eq{gklein} (e.g. the state in \Fig{kleings}),
though not all.
The smallest candidate for the cell $\cC$
on which to base the Hamiltonian \Eq{gklein}
such that there are no non-valence bond
ground states is the 12-site star-shaped 
cell depicted in Fig. (\ref{cells}a).
A detailed study of 
the uniqueness of the valence bond ground states
for this choice of cell
is left for future work. 
In the following,
the ideas discussed in this section will
be further generalized to allow the
construction of a Hamiltonian whose
only ground states, at least within the NN
valence bond basis, are the 
SRK-type RVB wavefunctions.

\subsection{The RVB Hamiltonian\label{RVB}}

While the generalized Klein models constructed
in the preceding subsection will have a lower
ground state degeneracy than the original Klein model \Eq{Hkl},
this degeneracy is still extensive. At the very least,
every valence bond state corresponding to an arbitrary
dimer covering $D$ of the lattice will be a ground state
of any generalized Klein model. 
In this Section, a local Hamiltonian is constructed
with ``resonating valence bond'' (RVB) type ground states,
which are certain superpositions of valence bond states. 
One desires these RVB ground states to be unique
and to describe quantum spin liquids with no spontaneously
broken symmetry. A proof of the uniqueness within the
restricted subspace of valence bond states will be given
in the following, whereas the generalization of the proof
to the full Hilbert space will rely on properties of the
generalized Klein models. 
%While these properties
%will not be proven here, they seem at least very likely,
%as discussed in the preceding section.
In order to have a fair amount of confidence
that the ground states describe spin liquids,
the Hamiltonian will be designed such that 
its ground state wavefunctions are spin-$\frac 12$
realizations of the Rokhsar-Kivelson  point 
of the quantum dimer model of the on the kagom\ea
lattice.\cite{misguich} These wavefunctions are in some sense the ``prototypical''
RVB-spin-liquid states, and at least for the quantum dimer model
describe a $\mathbb{Z}_2$ quantum liquid.\cite{misguich}
Whether or not this is still the case when dimers
are replaced by singlet valence bonds
is a non-trivial
question. It seems, however, likely that the answer is affirmative.
The main technical difficulty in answering this question
is the mismatch of the scalar product
 for corresponding states in the quantum dimer
and the spin-$\frac 12$ Hilbert space.  
Further discussion of this issue will be given in Section \ref{discussion}.

\subsubsection{The quantum dimer model on the kagom\ea lattice}

In a seminal paper, Moessner and Sondhi showed that 
the Rokhsar-Kivelson  point of the quantum dimer model (QDM)
on the triangular lattice has ground states describing
a $\mathbb{Z}_2$ topological quantum liquid. \cite{MS}
A subsequent work by Misguich et al.\cite{misguich}
generalized these findings to the kagom\ea lattice,
which was found to have several additional attractive 
features. In the present context, the most important distinctive feature
of the QDM on the kagom\ea lattice is the fact that
its RK-point lies in the interior of the $\mathbb{Z}_2$-liquid
phase. This is in contrast to the triangular case, where the RK-point
of the QDM lies at a (apparently first order) phase boundary.
This will guarantee the uniqueness of the four ``liquid'' ground states
of the model constructed here within the valence bond subspace,
and probably beyond, as argued in Section \ref{gen2}.

\begin{figure}[t]
\centering
\includegraphics[width=8cm]{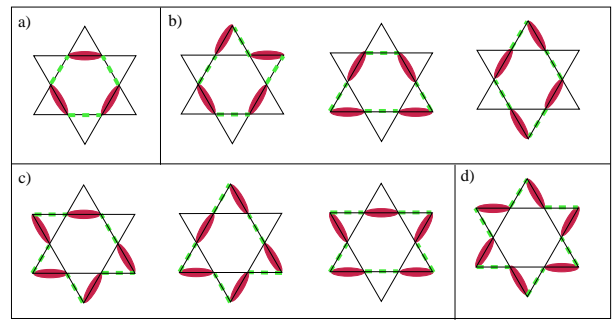}
\caption{ 
\label{loops}
Eight different types of dimer loops around a central hexagon.\cite{elser95}
Loops may be formed by three (a), four (b), five (c), or six (d) 
dimers. Each loop can be realized by two different dimer configurations
related by a resonance move. Dashed lines indicate dimer configurations
after a resonance move. The loops should be regarded as the transition graphs
between the original and the resonated configuration, i.e. the set of all
links carrying either a dimer or a dashed line. Using rotational symmetry,
there are 32 different loops corresponding to 64 dimer configurations.
 }
\end{figure}

The Hamiltonian of the QDM on the kagom\ea lattice is a sum of operators
acting on star-shaped 12-site cells as depicted in Fig. \ref{cells}a).
In the following, the term ``12-site cell'' will always refer to cells
of this topology.
Any dimer covering of this cell, as defined in Section \ref{defs},
defines a loop\cite{elser95}  around the central hexagon, given by the (shortest)
line connecting all points touched by a dimer, see \Fig{loops}.
Each
loop, on the other hand, is associated with with two possible dimer coverings
of the 12-site cell. These two dimer coverings are related by a shift
of all dimers along the loop, exchanging links with and without a dimer.
Such a shift will be referred to as a ``resonance move'' along the
loop formed by the dimers. For definiteness, unless otherwise
noted I will now assume a
translationally invariant 
kagom\ea lattice with toroidal topology.
Every dimer covering $D$ of the lattice gives rise to a covering
$D_\cC$ of every 12-site cell $C$, and a corresponding loop.
The Hamiltonian of the QDM of this lattice is a sum of operators
acting on any 12-site cell of the lattice, where each
operator performs a resonance move on the loop of dimers
present on the cell it acts on. (See Ref. \onlinecite{misguich} for details.)
The ground states of this
Hamiltonian consist of equal\cite{note3} amplitude
superpositions of all dimer states within a topological sector.
Here, the term topological sector is used in a restricted sense, where
two dimer states belong to the same sector if they
can be transformed into
one another via resonance moves. On the kagom\ea lattice, however, there
seems\cite{misguich} to be no distinction between topological sectors
in this kinetic sense, and the four topological sectors on the torus
defined in terms of transition graphs or winding numbers
(for recent reviews, the reader is again referred to Refs.\onlinecite{ML, moessnerraman}).
The QDM on the kagom\ea lattice
thus has exactly four degenerate ground states at the RK point.
These were shown\cite{misguich, furukawa} to be quantum liquids and
argued to be in the $\mathbb{Z}_2$ universality class.

\subsubsection{Correspondence between dimer and valence bond states: Sign convention\label{sign}}

\begin{figure}[b]
\centering
\includegraphics[width=6cm]{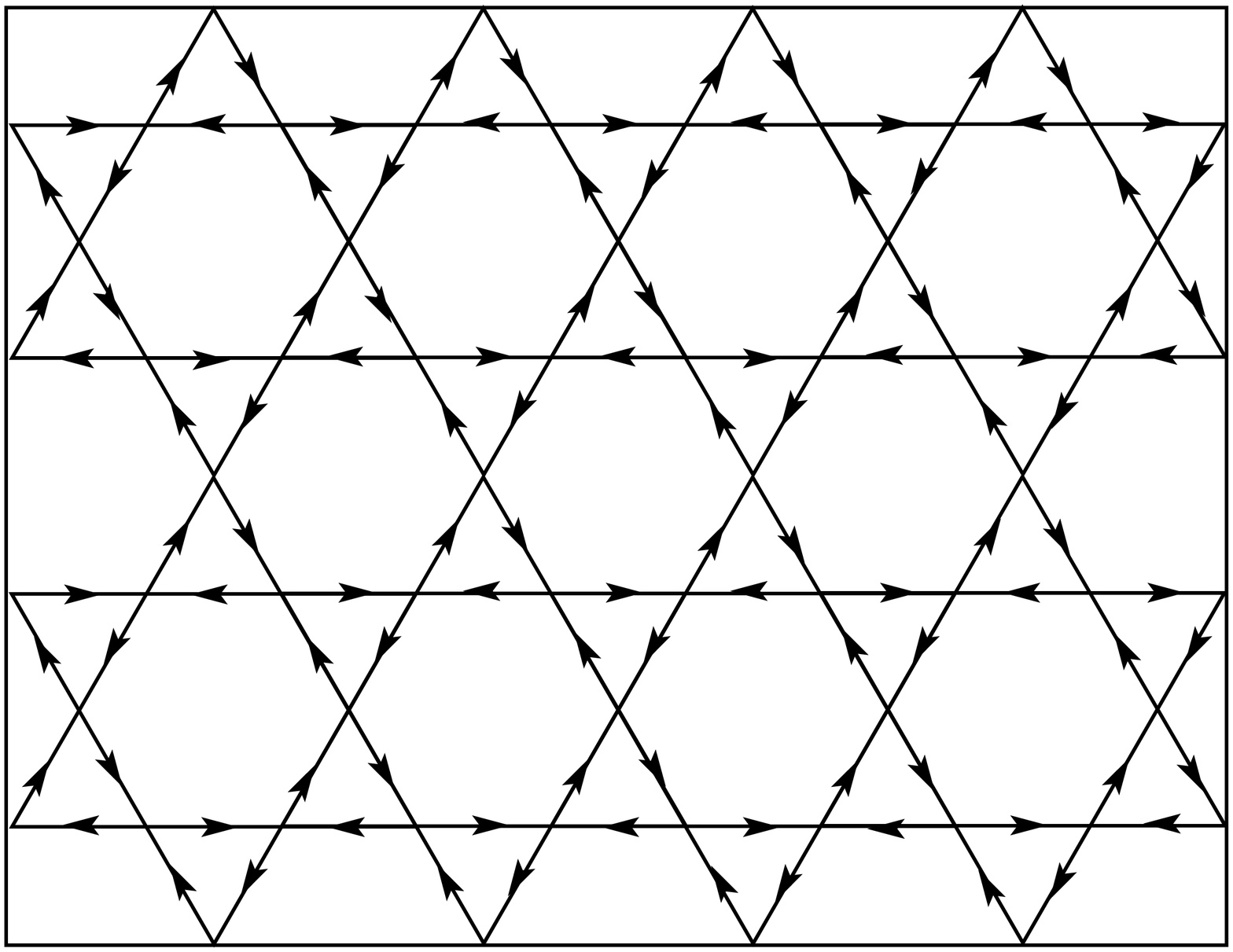}
\caption{ 
\label{orient}
Orientation of links on the kagom\ea lattice used to fix the sign 
of valence bond states. All links are oriented counterclockwise around 
the hexagon they belong to.
 }
\end{figure}

\begin{figure}[thb]
\centering
\includegraphics[width=2.5cm]{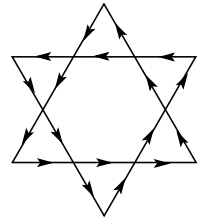}
\caption{ 
\label{orient2}
Preferred link orientation for the 12-site cell. 
All links are oriented counterclockwise around 
the central hexagon. This orientation is consistent
with the one shown in \Fig{orient} for the entire lattice,
as explained in the text.
 }
\end{figure}

It is easy to elevate the dimer liquid states just described
to states of superpositions of the spin-$\frac 12$ valence bond
states associated with each dimer basis state. As opposed to
the preceding Sections, the overall phases of valence bond states
now matter.  A convention for the overall phase of a valence bond 
state can be given by choosing an orientation for each link of the
lattice. This then fixes the sign of each valence bond singlet $[ij]=-[ji]$
on the  link $(i,j)$, and thus of the valence bond state, which is
a product of singlets. A suitable way to orient links is to do so
counter-clockwise around each hexagon, \Fig{orient}. 
A resonance move can now be 
viewed as a cyclic permutation of spins long a loop of dimers.
With the chosen orientation of links,
the sign associated with the state is preserved by such moves. 
To see this, note that for any given 12-site cell of the lattice,
flipping the orientation of the links touching boundary sites
does not change the sign of any valence bond state. This is so
since any of the possible dimer coverings of the 12-site cell shown in \Fig{loops} covers
an even number of such links. 
But with this new orientation, \Fig{orient2},
{\em all} links of the 12-site cell are oriented counter-clockwise
around the central hexagon. It is then clear that a cyclic permutation
of spins around a dimer loop preserves the overall sign associated
with the state.
These observations motivate that the orientation chosen in \Fig{orient}
is natural in the following sense: If the basis of valence bond states
is defined using the sign convention derived from this link orientation,
resonance moves on dimer states will translate into
cyclic permutations of spins in the associated
valence bond states. One might now be tempted to write
down an operator that performs such resonance moves
when acting on
valence bond states. 
This is possible because of the 
linear independence of the valence bond states.
Furthermore, due to the general results of Section \ref{li},
it is even possible to write such an
operator as a sum of local terms (acting, say,
on 19-site cells).
Such an operator cannot serve as a physical
spin-$\frac 12$ Hamiltonian, however, since it would not be Hermitian
with respect to the standard scalar product of the
spin-$\frac 12$ Hilbert space.
On the other hand,
these observations do not rule out the possibility
that one can construct
a local (Hermitian) Hamiltonian whose exact ground states are
just the equal amplitude superpositions of valence bond
states, within each topological sector and with the sign 
convention given here.
This question will be addressed in the following.

\subsubsection{Construction of the Hamiltonian\label{construct}}

I now consider a Hamiltonian of the following form
\begin{equation}
  \label{Hrvb}
H_{RVB}=-\sum_\cC R_\cC\;.
\end{equation}
where the sum goes over all 12-site cells (as defined above, Fig.\ref{cells}a) ) of the lattice, and the $R_\cC$ are certain Hermitian projection operators that enforce a ``resonance'' condition. For the time being, I will consider again a finite translationally invariant kagom\ea lattice $\cL$ of toroidal topology. Hence the lattice $\cL$ has no boundary sites.
The definition of the operators $R_\cC$ is as follows: For a given 12-site cell $\cC$, we first choose a sign convention for the 
valence bond states $\ket{D}$
in accordance with the link orientation in \Fig{orient2} . 
As explained the preceding Section, this convention is consistent with
the global sign convention chosen for valence bond states on the 
lattice $\cL$, \Fig{orient}. As discussed, every dimer pattern $D$ on $\cC$ corresponds
to one of two realizations of a certain loop around the central hexagon,
\Fig{loops}. By $D^\ast$ I now denote the other realization, related
to $D$ by a resonance move. 
In analogy with the definition of the set $\cB(\cC)$, \Eq{B},
we can now define a set of "resonant states" $\cR(\cC)$ via:
\begin{equation}\label{res}
\cR(\cC)=\{\,(\ket{D}+\ket{D^\ast})\,\otimes\ket{\psi_{D,j}}\,:D\in\cD(\cC), j=1\dotsc n_D\}.
\end{equation}
Form the elements of $\cR(\cC$), 
we can linearly generate any  state consisting of a resonant dimer loop,
with the free sites not touched by the loop in an arbitrary state. 
Note that $\free(D)=\free(D^\ast)$,
and we may without loss of generality assume that $\psi_{D,j}=\psi_{D^\ast,j}$. This will avoid
some unnecessary redundancy under the exchange of $D$ and $D^\ast$ in \Eq{res}.\cite{note4}
In complete analogy with the space $VB(\cC)$ of valence bond states on $\cC$,
one may now define the space $RL(\cC)$ of "resonance loop" states on $\cC$, via
\begin{equation}\label{RL}
  RL(\cC)=\sum_{D\in \cD(\cC)} \cH(\ket{D}+\ket{D^\ast})\;.
\end {equation}
The set $\cR(\cC)$ linearly generates the space $RL(\cC)$, 
and in fact turns out to be a basis 
of $RL(\cC)$, although this will not be used in the following. 
%This is in contrast to the set $\cB(\cC)$, whose elements are not
%linearly independent for the 12-site cell.
It is natural to define the operator $R_\cC$ to be the {\em orthogonal} projection
onto the subspace $RL(\cC)$. 
Note that the Hamiltonian \Eq{Hrvb} then has all the symmetries
of the underlying lattice. In addition, it is invariant
under  $SU(2)$ rotations, since the space $RL(\cC)$ 
is $SU(2)$-invariant for each $\cC$, for the same reasons
stated below \Eq{VB}.

Let us now consider Rokhsar-Kivelson-type spin-$\frac 12$ 
wavefunctions defined as follows,
\begin{equation}\label{RK}
  \ket{\psi}_\Omega=\sum_{D\in \Omega} \ket{D}\;,
\end{equation}
where $\Omega\subset \cD(\cC)$ contains all dimer coverings in a topological
sector as described above.
A similar type of state on the square lattice has been studied
early on by Sutherland.\cite{sutherland}
In view of this, I will refer to the spin-$\frac 12$ wavefunction \Eq{RK} as the "SRK"
state, to distinguish it from the RK-state of hardcore dimers.

It is quite easy to see that \Eq{RK} is indeed a ground state of 
\Eq{Hrvb} and, in fact, of every operator $-R_\cC$.
To see this, note that every state of the form 
\begin{equation}\label{rlstate}
(\ket{D_\cC}+\ket{D^\ast_{\cC}})\otimes\ket{S} \,,
\end{equation}
is invariant under the action of $R_\cC$, where again
$D_\cC$ is a dimer covering of the cell $\cC$, and $\ket{S}$
is any state on $\cL\backslash \supp(D_\cC)$. The reason
for this is entirely analogous to that given below
below \Eq{PD2}. 
\Eq{rlstate} can be thought of as having a
``resonance loop'' on the cell $\cC$.
Since the eigenvalues of $R_\cC$ are
0 and 1 by definition, \Eq{rlstate} is thus a ground state
of $-R_\cC$. We now write \Eq{RK} as a double sum
\begin{equation}
  \label{RK2}
    \ket{\psi}_\Omega=\sum_{D_\cC} \sum_{\overline{D_\cC}}\, \ket{D_\cC}\otimes\ket{\overline{D_\cC}}
\end{equation}
In here, the outer sum goes over all dimer coverings of $\cC$,
whereas the inner sum goes over all possible complements
$\overline{D_\cC}$ of $D_\cC$ such that $D_\cC\cup\overline{D_\cC}=:D$
is a dimer covering of $\cL$ in the topological sector $\Omega$. 
Note that for $D_\cC^\ast$ instead of  $D_\cC$, the possible 
choices for $\overline{D_\cC}$ are exactly the same, since
$D_\cC^\ast$ and $D_\cC$ have the same support, and the dimer
coverings
${D_\cC}\cup{\overline{D_\cC}}$ and 
${D_\cC^\ast}\cup{\overline{D_\cC}}$
are in the same topological sector by definition, since they differ
by a resonance move. We can thus also rewrite \Eq{RK2}
as
\begin{equation}
  \label{RK3}
    \ket{\psi}_\Omega=\sum_{(D_\cC,D_\cC^\ast)} \sum_{\overline{D_\cC}}\, (\ket{D_\cC}+\ket{D_\cC^\ast})\otimes\ket{\overline{D_\cC}}
\end{equation}
where the first sum now goes over (unordered) pairs $(D_\cC,D_\cC^\ast)$.
Since \Eq{RK3} is a sum over states of the form \Eq{rlstate},
it is manifestly invariant under the action of $R_\cC$.
This proves that $\ket{\psi}_\Omega$ is a ground state of $-R_\cC$
for each $\cC$. Hence $\ket{\psi}_\Omega$ is a ground state of the
Hamiltonian \Eq{Hrvb}.

\subsubsection{Proof of uniqueness within the valence bond basis.\label{uproof}}

So far we have succeeded in constructing a Hamiltonian
that has translational as well as $SU(2)$ invariance
and has ground states of the SRK form \Eq{RK}.
At the same time, one may hope that this Hamiltonian lacks
the extensive ground state degeneracy of the generalized Klein models
discussed in Section \ref{klein}. On the other hand, it is not yet clear
how many ground states \Eq{RK} has, both within as well as outside 
the valence bond subspace. In particular, since the construction
of $H_{RVB}$ 
closely follows that
of the generalized
Klein models, one may worry that some of the large degeneracy
of the latter remains in the present case. In this Section,
it will be shown that this is not so, in the sense that at least
{\em within} the valence bond state space $VB(\cL)$ there is only
one ground state of the form \Eq{RK} within each topological
sector $\Omega$, on a finite kagom\ea lattice $\cL$. 
In Section \ref{discussion}, it will be argued that   
either this already is the full degeneracy within the entire Hilbert space,
or one could find a perturbation such that this becomes the case.
For simplicity,
I will still assume periodic boundary conditions, and comment
on more general lattices in Section \ref{gen2}. 

Let us now consider a general state within the valence bond
subspace
\begin{equation}
  \label{trial}
\ket{\psi}=\sum_D a_D \ket{D}
\end{equation}
Suppose $\ket{\psi}$ is a ground state of $H_{RVB}$.
I will now show that the assumption that $\ket{\psi}$ is
{\em not}  a linear combination of the states displayed in \Eq{RK}
then leads to a contradiction.

\begin{figure}[t]
\includegraphics[width=7cm]{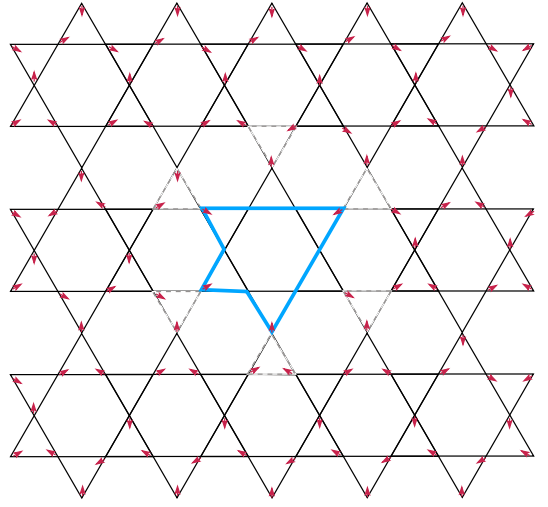}
\caption{\label{unique} Arrow representation for the dimer covering shown
in \Fig{kagome}, except for the internal hexagon of 
the central 12-site cell $\cC$. The arrows
shown are those that are uniquely determined by the condition that the
associated valence bond state survives the action of the operator $P$, \Eq{P2}.
The dashed triangles indicate links that belong neither to the cell $\cC$,
nor to any 19-site brick $\cC'$ that has no overlap with $\cC$.
The arrows at the corners of the dashed triangles are nonetheless all
determined (see text), and this determines the loop formed by the
dimer configuration on $\cC$ (fat line).
}
\end{figure}

The fact that \Eq{trial} is a ground state of $H_{RVB}$
implies that it is a ground state of each individual
operator $-R_\cC$. For, the SRK-states \Eq{RK} have this
property and would hence otherwise be lower in energy.
This then implies that $\ket{\psi}$ is invariant under the
action of each $R_\cC$,
\begin{equation}\label{inv}
R_\cC\ket{\psi}=\ket{\psi}\,.
\end{equation}
On the other hand, if $\ket{\psi}$ is not 
a linear combination of states of the form \Eq{RK},
there must be a pair of dimer coverings $D$ and $D'$
related by a single resonance move, such that $a_D\neq a_{D'}$.
For otherwise, it is easy to see that $\ket{\psi}$ would 
just be a superposition of SRK states, contrary to the assumption.
So there is a 12-site cell $\cC$ such that the dimer
loops contained in $\cC$ are $D_\cC$ and $D_\cC^\ast$
for $D$ and $D'$ respectively, whereas the 
remaining dimers are denoted by
$\overline{D_\cC}$ and are the same for $D$ and $D'$. We
thus have
\begin{equation}\label{DD}
  \begin{split}
    \ket{D}&=\ket{D_\cC}\otimes\ket{\overline{D_\cC}}\\
 \ket{D'}&=\ket{D_\cC^\ast}\otimes\ket{\overline{D_\cC}}\,.
  \end{split}
\end{equation}
Suppose, now, that we can find a projection operator $P$ which
commutes with $R_\cC$ and has the property that
\begin{equation}\label{P}
  P\ket{\psi}=a_D\ket{D}+a_{D'}\ket{D'}\,.
\end{equation}
Then, \Eq{inv} implies 
\begin{equation}\label{inv2}
  R_\cC\,(a_D\ket{D}+a_{D'}\ket{D'})=a_D\ket{D}+a_{D'}\ket{D'}
\end{equation}
By definition of $R_\cC$, we have
\begin{equation}
\begin{split}
  R_\cC (\ket{D}+\ket{D'})&=R_\cC\, (\ket{D_\cC}+\ket{D_\cC^\ast})\otimes\ket{\overline{D_\cC}}\\
&=\ket{D}+\ket{D'}
\end{split}
\end{equation}
where again the fact was used that $R_\cC$ leaves
states of the form \Eq{rlstate} invariant.
Multiplying the last equation by $(a_D+a_{D'})/2$,
subtracting from \Eq{inv2}, and dividing by
 $(a_D-a_{D'})/2$ (which is non-zero by assumption)
gives
\begin{equation}\label{inv3}
   R_\cC\,(\ket{D}-\ket{D'})=\ket{D}-\ket{D'}\,.
\end{equation}
By \Eq{DD}, the state on the right hand side of the last
equation is clearly of the general form
\begin{equation}\label{arlstate}
(\ket{D_\cC}-\ket{D^\ast_{\cC}})\otimes\ket{S} \,,
\end{equation}
with $\ket{S}$ some state on $\cL\backslash\supp(D_\cC)$ (here, $\ket{S}=\ket{\overline{D_\cC}}$).
\Eq{arlstate} is the counterpart of \Eq{rlstate}
with the resonance loop replaced by an ``anti-resonance loop''.
The latter are related to ``vison excitations'' in the
QDM on the kagom\ea lattice.\cite{misguich}
Two things now remain to be shown: 1.) that an operator
$P$ with the desired properties can be found, and 2.)
that no state of the form \Eq{arlstate} can be invariant under
the action of $R_\cC$, such that \Eq{inv3} leads to a
contradiction.
The second statement would easily follow if the
set $\cB(\cC)$ were linearly independent already 
for the 12-site cell. This is not so, and to avoid
technicalities of this nature here, the proof is deferred
to the Appendix.
%Nonetheless,
%it is in principle easy to verify this statement 
%numerically, since it can be reduced to a property of a
%finite cluster.
%An algebraic proof is, however, given in Appendix \ref{}.
Here I will focus on the construction of the operator $P$, \Eq{P}.
Consider
\begin{equation}\label{P2}
  P:=\prod_{\cC'} \,P_{D_{\cC'}}  \,,
\end{equation}
where the product goes over all 19-site 
bricks $\cC'$ that  
have no sites in common with the fixed 12-site cell  $\cC$ within which $D$ differs from $D'$. 
The projection operators $P_{D_{\cC'}}$ are those defined in
\Eq{PD0}, and $D_{\cC'}$ is the restriction of $D$ onto $\cC'$ as always.
It is then clear that $R_\cC$ commutes with $P$, since
$R_\cC$ only acts on $\cC$ whereas $P$ acts only on the complement $\overline \cC$ of $\cC$ in $\cL$. Furthermore, it is clear that the action of $P$ leaves the state $\ket{D}$ invariant by construction (cf. \Eq{PD3}) , and the same is true for the state $\ket{D'}$, since $D$ and $D'$ do not differ on $\overline \cC$.
It remains to show that $P$ annihilates every valence bond state
other than $\ket{D}$ and $\ket{D'}$.
To this end, it is best to introduce the arrow representation for 
dimer coverings\cite{elser_arrow}, as shown in \Fig{unique}.
Here, one assigns an arrow to each lattice site, which points to
the center of either of the adjacent triangles. These arrows are 
further subject to the constraint that each triangle must have either  two
inward pointing arrows or none. We associate
a dimer with any link between two inward pointing arrows on any
given triangle. It is easy to see that the allowed arrow states are in 
one-to-one correspondence with the dimer coverings of $\cL$.
Furthermore, the knowledge that a valence bond state
$\ket{D''}$ survives the action of the operator $P_{D_{\cC'}}$
already determines the arrows associated with $D''$ for all sites in $\cC'$ :
All dimers of $D''$ that are fully contained in
$\cC'$ must be identical to 
dimers in  $D$, i.e. $D''_{\cC'}=D_{\cC'}$. Hence the arrows on sites touched
by such dimers are determined, while those of the remaining
boundary sites of $\cC'$ must point outward, i.e. away from $\cC'$. 
Note that the latter sites 
just make up the set $\free(D_{\cC'},\cC')$.
Thus if $\ket{D''}$ survives the action of $P$, the
arrows corresponding to $D''$ are determined 
for all sites in $\overline \cC$.
This follows since any such site belongs
to a 19-site brick $\cC'$ that has no overlap with $\cC$ (cf. \Fig{unique}),
provided that the lattice is sufficiently large (in both directions), which
will be assumed in the following.
%Note that this is so even though not all {\em links}
%contained in $\overline \cC$ belong to such a $\cC'$, \Fig{unique}.
Furthermore, as is apparent from \Fig{unique}, this also
determines the arrows of $D''$ on boundary sites of $\cC$,
 since when two arrows of a triangle are
determined, then so is the third. % (cf. \Fig{unique}). 
Hence, the fact that 
$\ket{D''}$ survives the action of $P$ determines all
the arrows associated with $D''$ {\em except} those on the interior
sites of $\cC$. The arrows thus determined must be identical
to those of $D$ and $D'$, since the corresponding valence
bond states $\ket{D}$ and $\ket{D'}$  likewise survive the action of $P$. 
However, as mentioned above, the arrows on the 
boundary sites of $\cC$ determine the set $\free(D''_{\cC},\cC)$
and thus determine the loop-type associated with $D''_{\cC}$.
The remaining choices for the arrows on the internal sites
of $\cC$ then correspond to the two possible realizations of
this loop, which then precisely lead to the dimer coverings
$D$ and $D'$. Hence, $D''$ must be equal to either $D$ 
or $D''$.
 This concludes the proof of the two properties
required of $P$, namely $[R_\cC,P]=0$ and \Eq{P}.
Note that the construction of the operator $P$
is essentially possible  due to the linear independence
of the valence bond states on $\overline C$, even
though $\overline C$ is not regular. (All sites
of $\overline C$ belong to a 19-site brick, but
not all links of $\overline C$. Cf. \Fig{unique} and the discussion in
Section \ref{gen1}.)
 I state the results of this section as

{\em Theorem III.} Let $\cL$ be a sufficiently large finite periodic 
kagom\ea lattice. Then the  SRK states \Eq{RK}
and their linear combinations are the only ground states of
the Hamiltonian $H_{RVB}$, \Eq{Hrvb},
 within the valence bond subspace  $VB(\cL)$.

\subsubsection{Possible generalizations of Theorem III.\label{gen2}}

The most important generalization of Theorem III one can consider
relates to the obvious question whether or not the uniqueness of the SRK
ground states holds within the entire Hilbert space.
Here I will briefly show that this question is naturally analyzed 
in two steps, the first of which is just the uniqueness within the valence bond
subspace, as stated in Theorem III. The second step, which is the
extension of this statement to the full Hilbert space, may likely be 
analyzed in terms of  the (arguably simpler)
generalized Klein model \Eq{gklein} for the 12-site cells, now denoted  $H_{GK,12}$. 
More precisely,
the following implication holds: If it can be shown that the generalized Klein
model $H_{GK,12}$ has no ground states outside
the NNVB subspace, then this is also true
for $H_{RVB}$. 
This follows from the fact that every ground state of $H_{RVB}$
is necessarily also a ground state of $H_{GK,12}$. 
To see this, it is best to focus on the local density
matrix $\rho_{\cC}$ of a 12-site cell $\cC$ for a ground state
$\ket{\psi}$ of $H_{RVB}$. 
As discussed above, $\ket{\psi}$ is in particular a ground state
of $-\cR_\cC$.  The ensemble described by $\rho_{\cC}$
can thus have no weight outside the space $RL(\cC)$, \Eq{RL}.
But since $RL(\cC)\subset VB(\cC)$, such a state is also a ground state of
$P_\cC$, defined above \Eq{gklein}. Since $\cC$ was arbitrary, $\ket{\psi}$
is thus a ground state of $H_{GK,12}$.
Needless to say, the reverse implication need not hold.
Hence even if the generalized Klein model $H_{GK,12}$ 
has non-valence-bond ground states, this need not
be true for $H_{RVB}$. Moreover, if {\em both} $H_{RVB}$ and $H_{GK,12}$ had
non-valence-bond ground states, it may be possible to
find a generalized Klein model based on a larger cell $\cC'$
(e.g. the 19-site bricks) which does not have such ground states.
In this case, one can render the SRK  ground states unique
by adding this generalized Klein Hamiltonian (multiplied
by an arbitrary positive constant) to $H_{RVB}$. 
However, it seems likely that $H_{GK,12}$ already
has the desired properties, in which case no such perturbation
is needed.  I note that for the usual Klein models,
it seems that in all well studied cases,
the NNVB ground states are either unique\cite{kcc},
or there are obvious (i.e. simple product state) exceptions,
such as the one shown in \Fig{kleings}.
The latter does not seem to be the case for $H_{GK,12}$.
One may thus be hopeful that the uniqueness of the NN
valence bond ground states can be explicitly demonstrated in this case,
just as it has been possible for certain Klein models\cite{kcc} (cf. also Ref. \onlinecite{nussinov2}).
A systematic study of this problem is reserved for future work.

Another natural generalization to consider is the application
of the present construction to different lattices. For the square and
honeycomb lattice, 
a different method is already available.\cite{fujimoto}
However, the method discussed here is also applicable in these cases, see below.
The solvable points 
of the underlying quantum dimer models
on these bipartite lattices correspond to 
critical states. This may also be true 
of the corresponding spin-$\frac 12$ Hamiltonians.
More akin to the kagom\ea case discussed here is the
triangular lattice, at least from the point of view of 
quantum dimer models.\cite{MS}
In the triangular case, the RK-point also describes a $\mathbb{Z}_2$
liquid phase, albeit at a first order phase boundary.
Furthermore, the RK-points of quantum dimer models on 
bipartite lattices in three dimensions
have been argued to describe stable critical phases.\cite{huse,MS2,hermele}
It seems desirable to generalize the methods developed here to all these cases.

On any lattice, the following generalization of the
present construction suggests itself (cf. again Ref. \onlinecite{note5}).
We consider a cell $\cC$ which is sufficiently large.
This implies that
all types of resonance moves of the quantum dimer model
on the same lattice may take place within cells of this kind,
and also that the cell has non-vanishing interior.
We consider 
SRK-type
%the spin-$\frac 12$ versions of the RK-
wavefunctions,
and ask whether for these states, the resulting density matrix $\rho_\cC$ of that cell
is restricted to a certain subspace of $\cH(\cC)$. 
The answer will in general be affirmative, for large enough $\cC$. 
In fact, $\rho_\cC$
will be restricted to (i.e. have no weight outside of)
a certain subspace $RL(\cC)\subset VB(\cC)$.
Here, $VB(\cC)$ is constructed just as before from
all possible restrictions of dimer coverings
to the
cell $\cC$, with free sites in an arbitrary state.
$VB(\cC)$ may be used to define generalized Klein models,
as discussed in Section \ref{gen1}.
To construct $RL(\cC)$, we introduce equivalence
classes on the set $\cD(\cC)$ of dimer coverings of
$\cC$, where two coverings belong to the same class
if they are related by a series of resonance moves
taking place {\em within the cell} $\cC$. 
In analogy with \Eq{res}, the set $\cR(\cC)$ of resonating states
on $\cC$ is then defined as equal amplitude superposition
of valence bond states within one equivalence class.
Again, the state of the free sites is chosen from an arbitrary basis
(but is fixed for any such
superposition). It is then found that the space $RL(\cC)$
spanned by all states in $\cR(\cC)$ contains all the
non-zero weight of the local density matrix $\rho_\cC$
in an SRK-type state. These states are thus ground states
of a Hamiltonian constructed in analogy with
\Eq{Hrvb}.  It is clear that these observations 
are completely analogous to those
made for the kagom\ea lattice above,
except that a language free of density matrices
has been given preference there.
Likewise, a discussion of the uniqueness of the SRK ground states
on general lattices
should in most cases be feasible along lines similar to those for the kagom\ea case
discussed above.

Finally, it is expected that the findings of the preceding sections
are not limited to periodic lattices, but can be carried
over to reasonably benign lattices with an edge.
On any such lattice, one would want to add a prescription
for boundary terms of the Hamiltonian \Eq{Hrvb}. 
For, in the absence of translational symmetry, there
is no reason why these should be identical to the
bulk terms already defined. In fact, for 12-site
cells $\cC$ that lie at the boundary of the lattice,
one would again want to alter the definition of the
set $\cR(\cC)$, \Eq{res}, by discarding all dimer coverings
$D\in\cD(\cC)$  where some boundary sites of $\cL$ are not
touched by a dimer. Otherwise, there would be gapless edge 
excitations, where spins at boundary sites are put into an
arbitrary state and do not participate in valence bonds.
Such gapless edge modes are not generically present in a 
$\mathbb{Z}_2$ topological state. However, the fact that 
the above modification easily gets rid of these modes
indicates that a generic edge perturbation would do the same.
A merely technical subtlety arises in the construction
of the operator $P$, \Eq{P2},  in the proof of Theorem III.
When a 12-site cell $\cC$ is near the edge, it may not be possible
to cover its complement $\overline \cC$ by 19-site bricks.
One may, however, find other ``linear independence bricks''
near the edge, which may be smaller than the 19-site bricks,
since for edge cells $\cC'$  the set $\cB(\cC')$  would be subject
to the same truncation discussed above for the set $\cR(\cC)$.

\section{Discussion of the physical implications of the solvable point\label{discussion}}

In the above, a point in the phase diagram of local $SU(2)$-invariant spin-$\frac 12$ Hamiltonians
on the kagom\ea lattice has been identified
for which
exact ground states can be found.
A proof of the uniqueness of these ground states
within a restricted Hilbert space has been given.
It has been argued that this uniqueness likely
holds within the full Hilbert space, possibly
involving slight modifications of the Hamiltonian,
and strategies have been outlined to prove this.
%Furthermore, a strategy has been outlined
%that may lead to the extension of this proof
%to the entire Hilbert space,  possibly involving slight modifications
%of the Hamiltonian.
Some interesting questions remain to be resolved by future work,
which will be addressed only at a qualitative level here.
These questions are: ``Is there a gap between the ground state sector and the excited states?'' ``How do correlations behave at long distances?''
``Do the ground states break any symmetry ?''
These questions are intimately related,  both to each other and to the question of how much of the properties of the kagom\ea lattice QDM survives in the present realization of the RK-point through spin-1/2 degrees of freedom. 
One may certainly hope that correlations largely carry over from the quantum dimer case, where it has been shown that any correlations between local operators
are short ranged.\cite{furukawa} The main difficulty in generalizing this proof to the present case is the non-orthogonality of the valence bond states. It seems that this may not change the physics much, since the overlap between two random valence bond states tends to be very small, and is only appreciable for similar valence bond configurations. 
%This is in agreement with the findings of Ref. \cite{somebody}. 
(See, however, Ref. \onlinecite{fendley2} for a discussion of the
impact of the 
choice of
scalar product in a related but different problem.)
The expected behavior is thus that the ultra-short ranged correlations of the QDM become exponentially decaying for the  SRK-states  \Eq{RK}. If so, this would preclude the existence of broken symmetry, and would be a strong argument in favor of a gap. One may caution that
special Hamiltonians are known which are gapless despite short ranged correlations.\cite{freedman}
However, this is not generically expected, and the knowledge
that on any finite torus
the ground state precisely has a four fold topological degeneracy,
combined with the absence of symmetry breaking, would be quite compelling evidence that the low temperature phase of \Eq{Hrvb} is the same as that of the kagom\ea lattice QDM. The latter is known to be a gapped $\mathbb{Z}_2$-liquid.\cite{misguich} 
Establishing the existence of an energy gap directly will likely require numerical efforts. 
On the other hand, it seems possible that the particularly benign properties 
of the kagom\ea lattice, which have given rise to strong exact statements about dimer correlation functions for its RK-states,\cite{furukawa} may allow insights into correlations in the present case as well. 
This would require one to tackle the issue of non-orthogonality, as discussed above, 
and will be left for future work.

I note that one attractive feature of the kagom\ea lattice quantum dimer model is the exact knowledge of all eigenstates,\cite{misguich} including spinons\cite{KRS} and Ising vortex\cite{read,kivelson} excitations,
called ``visons'' in the recent literature.\cite{SF}
This does not carry over to the present case, as the associated spin-$\frac 12$ wavefunctions
would not be eigenstates of the Hamiltonian \Eq{Hrvb} for any obvious reasons.
They would, however, be natural variational candidates. Note that on the other hand,
these excitations do not disperse for the solvable quantum dimer model, 
but should do so in the present case, which is certainly the generic behavior
(cf., e.g., Ref. \onlinecite{misguich2}).

The solvable Hamiltonian constructed here seems somewhat unrealistic, due to the presence of operators that act on 12 spins at a time. It is not immediately clear how dominant such terms are when the local operator $R_\cC$ is expanded in 
two-spin
and higher order exchange terms. While explicitly carrying out such an expansion would be worthwhile, there is no reason to assume that nearest neighbor two-body processes will dominate. Even so, 
there is much to be said in favor of the usefulness of an exactly solvable ``reference point''. Firstly, if the solvable point turns out to be gapped, as is expected in the present case, it must lie within the interior of a phase. This phase will survive at least small perturbations in the direction of more realistic Hamiltonians, and there is a distinct possibility that one may make these perturbations large enough to reach a realistic regime without encountering a phase boundary. Secondly, while establishing the existence of a liquid phase numerically is exceedingly difficult due to size limitations, it may be somewhat less so to establish that two points belong to the same phase, especially if this phase has a robust gap along a line connecting these points. Hence, to establish the properties of a single reference point may be of considerable benefit, even if the reference point itself is unphysical. Thirdly, 
one may expect that higher order exchange terms, as certainly present in \Eq{Hrvb}, can be of considerable importance on the insulating side near a metal-insulator transition, where a Hubbard-type expansion parameter $t/U$ is not small. Arguments of this type have been made,\cite{lesik} at least for 
four-spin ring exchange terms,  in the concrete example of the triangular antiferromagnet
$\kappa$-(ET)$_2$Cu$_2$(CN)$_3$.\cite{triangular1}

It is worth noting that
the recently studied "herbertsmithite" kagom\ea antiferromagnet\cite{kagome2,kagome3,kagome4} may fit into
a picture based on a gapless "Dirac" spin liquid state.\cite{hastings2,ran,hermele2}
The gapless nature of this system is indicated by various
experimental probes, such as low temperature
susceptibility and specific heat. Magnetic impurities
may however play an important role in these low
temperature properties (see, e.g., Ref. \onlinecite{PALee}
for a brief summary of results).
The presence of gapless excitations is also supported by some recent
numerical studies,\cite{sheng} though only for the singlet sector.
A physical picture for gapless singlet excitations has been offered in
Ref. \onlinecite{Hao}.

If the nature of the low lying excitations of the herbertsmithite compound is gapless, the low temperature phase of this system does not seem to be directly
related to the solvable point described in this work.
Nonetheless, the existence of a solvable point
of this kind may open up the possibility that a topological spin liquid state 
can in principle
be realized
in $SU(2)$-invariant kagom\ea antiferromagnets, 
if there is some mechanism that generates 
sufficiently high order spin couplings.
I note that a very similar conclusion has been reached before
in Ref. \onlinecite{ashvin} based on  a projective symmetry 
group\cite{wenpsg} analysis of Schwinger boson states.

%To determine how large these couplings need to be
%in practice will require detailed numerical studies.

\section{Conclusion\label{conclusion}}

In this work, the linear independence of
nearest neighbor valence bond states on the kagom\ea
lattice has been proven, using a method that may
allow generalization to other lattices as well.
Furthermore, capitalizing on techniques used in the proof,
a class of spin-$\frac 12$ model Hamiltonians 
has been constructed whose ground states
simultaneously minimize the energy of non-commuting local projection
operators. 
%The focus has been on kagom\ea type lattices.
One variant of these projection operators leads to the
notion of "generalized Klein models''.
Another variant of these operators,
with a more restricted image, leads to an $SU(2)$-invariant 
local Hamiltonian whose ground states within the
nearest neighbor valence bond manifold are uniquely given
by the four topologically degenerate "Sutherland-Kivelson-Rokhsar" states
on toroidal kagom\ea lattices. 
%This is in contrast to the
%(generalized) Klein models, whose ground state degeneracy within the same
%manifold is extensive. 
It is argued that these ground states describe a
$\mathbb{Z}_2$ topological quantum liquid with unbroken
translational and $SU(2)$-rotational invariance. 
This is based on the
close analogy to similar
ground states of a quantum dimer model on the same lattice, where
a notion of rotational invariance is %, however,
 lacking.
Questions pertaining to the uniqueness of the SRK
ground states within the full Hilbert space have been reduced to properties
of the generalized Klein Hamiltonians, whose detailed study is left to
future work.
I am hopeful that followup work on the remaining questions
raised in this paper will establish the existence of the
$\mathbb{Z}_2$ topological phase within the phase
diagram of $SU(2)$-invariant spin-$\frac 12$ Hamiltonians
on the kagom\ea lattice, and possibly other lattices, beyond reasonable doubt.

\begin{acknowledgments}
I am indebted to Z. Nussinov and E. Fradkin for stimulating discussions.
This work was supported in part by the National Science Foundation under Grant No. PHY05-51164 during a stay at the Kavli Institue for Theoretical Physics,
and under NSF Grant No. DMR-0907793.

\end{acknowledgments}

\appendix

\section{Exclusion of vison configurations from the ground state manifold}

In this Appendix, a technical lemma is derived which completes the proof of the
uniqueness of the SRK ground states within  the NNVB
subspace, Theorem III.
The proof of this theorem was based in part on the following observation:
Consider a 12-site cell $\cC$ of the topology shown in Fig. \ref{cells}a).
On this cell we consider a state of the general form \Eq{arlstate},
which I restate here as
\begin{equation}\label{vison}
(\ket{D}-\ket{D^\ast})\otimes\ket{S} \,.
\end{equation}
Here, $D$ is a dimer covering of the 12-site cell, $D^\ast$ is 
its counterpart related to $D$ by a resonance move. $\ket{S}$
is some state on the remaining sites of the lattice,
$\cL\backslash\supp(D)$. Then, the statement to be shown
is that no state of the form \Eq{vison} may be invariant
under the action of the operator $R_\cC$ constructed in Section \ref{construct}.
The first step is to recast this statement as a property of the 12-site cell alone,
disentangled from the remaining lattice. To this end,
one may observe that it is sufficient to show that no state of the 
following form,
\begin{equation}\label{vison2}
(\ket{D}-\ket{D^\ast})\otimes\ket{\psi} \,,
\end{equation}
may be a ground state of the operator  $-R_\cC$, where $\ket{\psi}$
is a state on $\free(D,\cC)$. \Eq{vison2} is thus a state on $\cC$.
One way to see this is to observe that in the state \Eq{vison},
the local density matrix $\rho_\cC$ for the 12-site cell can be written
as a sum of orthogonal projection operators onto states of the
form \Eq{vison2}. It then follows that if no state of the form \Eq{vison2} 
can be a ground state of $-R_\cC$, then neither can any state
of the form \Eq{vison}. The latter is equivalent to the fact that
no state of the form \Eq{vison} may be invariant under the action of $R_\cC$.
\begin{figure}[t]
\centering
\includegraphics[width=8cm]{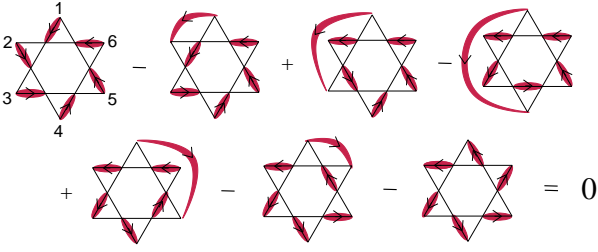}
\caption{ 
\label{lindep}
One of six linear dependences of the states in $\cB(\cC)$ for the 12-site cell.
The remaining five are related to the one shown here by rotation.
See text and \Eq{6lindep}. }
\end{figure}

All that remains to be shown is thus that no state of the form \Eq{vison2}
exists in the linear span of $\cR(\cC)$, \Eq{res}. Again, this would be trivial if
the set $\cB(\cC)$ consisted of linearly independent states, but this is
not the case for the 12-site cell considered here. Luckily, one finds that there
are only six linear relations among the 730 states of the set $\cB(\cC)$.
These involve only the singlet sector, and only 5- and 6-dimer loops.
The two 6-dimer loop states are automatically singlets,
whereas singlet 5-dimer loop states have two free sites that form a singlet bond.
\Fig{lindep} graphically depicts one such linear relation, whereas the remaining
5 relations are obtained by rotating the diagrams in \Fig{lindep}.
In this figure, each bond denotes a singlet, with orientations
indicated by arrows . The validity of the identity shown in the
figure follows easily from the following graphical identity for any 
pair of singlet bonds between four spins:
\begin{equation}
 \includegraphics[width=7cm]{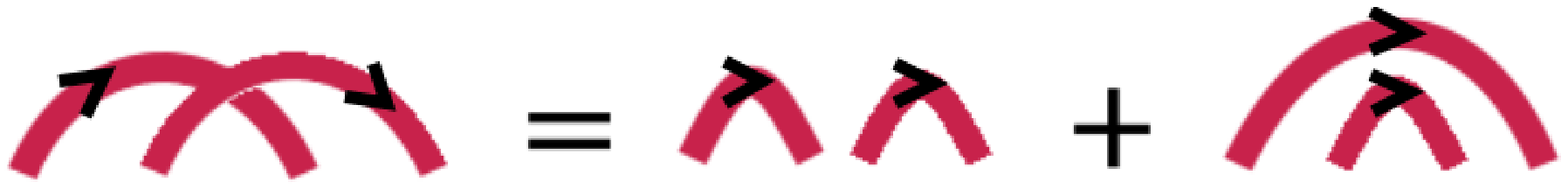}\,.
\end{equation}
We may also write the six linear dependences in a more compact form
as follows:
\begin{equation}\label{6lindep}
 \begin{split}
   6L-(1,2)+(1,3)-(1,4)+(1,5)-(1,6)-6L^\ast =0\\
   6L-(2,3)+(2,4)-(2,5)+(2,6)-(2,1)-6L^\ast =0\\
   6L-(3,4)+(3,5)-(3,6)+(3,1)-(3,2)-6L^\ast =0\\
   6L-(4,5)+(4,6)-(4,1)+(4,2)-(4,3)-6L^\ast =0\\
   6L-(5,6)+(5,1)-(5,2)+(5,3)-(5,4)-6L^\ast =0\\
   6L-(6,1)+(6,2)-(6,3)+(6,4)-(6,5)-6L^\ast =0
 \end{split}
\end{equation}
Here, $6L$ denotes the first 6-dimer loop state in \Fig{lindep},
$6L^\ast$ the last. A term $(a,b)$ denotes a 5-dimer loops state
with free sites $a$ and $b$ joined by a singlet bond from $a$ to $b$,
where the boundary sites of $\cC$ are labeled as in the first graph of the figure. Here, $b$ denotes
the free site with a dimer on the same triangle, and $a$ denotes the free
site without a dimer on the same triangle. This determines all the remaining 
dimers, and nearest neighbor valence bonds are by definition
oriented counter-clockwise around the central hexagon.
The first line in \Eq{6lindep} is thus exactly the relation depicted in \Fig{lindep}.
Note that each $(a,b)$ appears in one and only one line of \Eq{6lindep}.
We may further note
\begin{equation}\label{dual}
(a,b)=-(b,a)^\ast\,.
\end{equation}
Here, $(a,b)^\ast$ denotes the state obtained from $(a,b)$ by
shifting all nearest neighbor valence bonds along the loop
they form, but leaving the singlet between sites $a$ and $b$
untouched (hence the overall minus sign, since the singlet
bond between $a$ and $b$
has opposite orientations in $(a,b)$ and $(b,a)$).
The fact that the relations in \Eq{6lindep} are the {\em only}
linear relations between the states of $\cB(\cC)$
can be shown analytically\cite{ASU} by using Rumer-Pauling
valence bond diagrams \cite{rumer, pauling, soos1,soos2}. In addition, it is easy to verify
this fact numerically, which I have carried out using Ref. \onlinecite{LinBox}.
Let us now assume that a state of the form \Eq{vison2}
is contained in $RL(\cC)$, i.e. in the linear span of $\cR(\cC)$.
That is, we assume that there is a relation of the form
\begin{equation}
  \label{span}
(\ket{D}-\ket{D^\ast})\otimes\ket{\psi} =\sideset{}{'}\sum_{D',j}  \lambda_{D',j}(\ket{D'}+\ket{D'^\ast})\otimes\ket{\psi_{D',j}}
\end{equation}
and want to show that this leads to a contradiction.
Here, the prime restricts the sum to one of the two dimer coverings per loop.
One may first observe that the dimer covering $D$ on the left
hand side must correspond to a 5- or 6-dimer loop.
Otherwise, there would be a non-trivial linear relation involving a
3- or 4-dimer loop valence bond state, and such a relation
does not exist, since \Eq{6lindep} is a complete set of linear
relations. Furthermore, the state in \Eq{span} cannot survive
a projection onto the subspace of non-zero total spin (which affects
only the $\ket{\psi}$ factors).
For otherwise, there would be a non-trivial linear relation
involving non-singlets, which again does not exist.
Hence, \Eq{span} must be of one of the following two forms:
\begin{subequations}
\begin{align}
6L-6L^\ast&=\lambda_6(6L+6L^\ast)+\sum_{a'<b'} \lambda_{a',b'}[(a',b')-(b',a')]\label{lc1}\\
(a,b)+(b,a)&=\lambda_6(6L+6L^\ast)+\sum_{a'<b'} \lambda_{a',b'}[(a',b')-(b',a')]\label{lc2}
\end{align}
\end{subequations}
where \Eq{dual} was taken into account.
Let us focus on \Eq{lc1} first. As a non-trivial
linear relation, it must be possible to obtain \Eq{lc1} as
a linear combination of the 6 relations in \Eq{6lindep},
\begin{equation}
  \label{lc3}
  \sum_{i=1}^6 \mu_i \ell_i\,,
\end{equation}
where the $\ell_i$ represent the six lines of \Eq{6lindep}.
The requirement that $(a',b')$ and $(b',a')$ must enter
with opposite signs results in the requirement that $\mu_i=-\mu_j$
for any $i\neq j$, which is evidently impossible.
Next, let us try to obtain \Eq{lc2} from \Eq{lc3}. 
The situation is similar. There is now a single pair
$(i,j)=(a,b)$ for which the relation $\mu_i=-\mu_j$
need not hold. For any $i\neq j$ with $(i,j)\neq (a,b)\neq (j,i)$,
$\mu_i=-\mu_j$ still follows from the same reasoning as before.
Still, this is impossible to satisfy. This concludes the proof
that no state of the form \Eq{vison2} can be expressed
as a linear combination of the ``resonance loop'' states
making up the set $\cR(\cC)$, and hence no
state of the form \Eq{vison} is invariant under the action
of the operator $\cR_\cC$, as explained initially.

%In particular,
%this also determines the set
%$\free({D''}_\cC,\cC)$, and hence
%the loop formed by the dimers in ${D''}_\cC$. 

%\bibliography{kagome}

\begin{thebibliography}{86}
\bibitem[{not({\natexlab{a}})}]{note1}
\bibinfo{note}{{{Some convention is necessary to define the action of $P_b$
  within the full Hilbert space $\cH(\cC)$, which amounts to choosing a direct
  complement of $VB(\cC)$ within $\cH(\cC)$. This is not of any importance in
  this Section} }}.
\bibitem[{not({\natexlab{b}})}]{note5}
\bibinfo{note}{On general lattices, one may define internal sites of cells
  again via maximal coordination number. Links between boundary sites are not
  to be included into cells. In other words, each dimer of a covering of a cell
  must touch at least one internal site, and restrictions of dimer coverings
  onto cells are to be defined accordingly.}
\bibitem[{not({\natexlab{d}})}]{note4}
\bibinfo{note}{That is, if $\psi_{D,j}\neq\psi_{D^\ast,j}$, more states would
  be included into the set $\cR(\cC)$ without affecting its linear span.}
\end{thebibliography}

\end{document}